\documentclass[pra,aps,amsmath,amssymb,notitlepage,twocolumn,floatfix,superscriptaddress,10pt]{revtex4-1}
\usepackage[bookmarks=false,colorlinks,citecolor=blue,urlcolor=blue]{hyperref}
\usepackage{amsthm}
\usepackage{amsfonts}
\usepackage{siunitx}
\usepackage{amsmath}
\usepackage{amssymb}
\usepackage{graphicx}
\usepackage{color}
\usepackage{textcomp}
\usepackage{xcolor}
\usepackage{helvet}
\usepackage[eulergreek]{sansmath}
\usepackage{afterpage}

\DeclareMathOperator*{\argmin}{argmin}
\DeclareMathOperator*{\arccot}{arccot}

\newlength\figureheight
\newlength\figurewidth

\newcommand{\fH}{\mathcal{H}}
\newcommand{\fC}{\mathcal{C}}

\renewcommand{\epsilon}{\varepsilon}

\begin{document}
\title{Degeneracy, degree, and heavy tails in quantum annealing}
\author{Andrew~D.~King}\email[]{aking@dwavesys.com}
\affiliation{D-Wave Systems Inc., Burnaby B.C.}
\author{Emile~Hoskinson}
\affiliation{D-Wave Systems Inc., Burnaby B.C.}
\author{Trevor~Lanting}
\affiliation{D-Wave Systems Inc., Burnaby B.C.}
\author{Evgeny~Andriyash}
\affiliation{D-Wave Systems Inc., Burnaby B.C.}
\author{Mohammad~H.~Amin}
\affiliation{D-Wave Systems Inc., Burnaby B.C.}
\affiliation{Department of Physics, Simon Fraser University, Burnaby B.C.}
\date{\today}

\begin{abstract}
Both simulated quantum annealing and physical quantum annealing have shown the emergence of ``heavy tails'' in their performance as optimizers: The total time needed to solve a set of random input instances is dominated by a small number of very hard instances.  Classical simulated annealing, in contrast, does not show such heavy tails.  Here we explore the origin of these heavy tails, which appear for inputs with high local degeneracy---large isoenergetic clusters of states in Hamming space.  This category includes the low-precision Chimera-structured problems studied in recent benchmarking work comparing the D-Wave Two quantum annealing processor with simulated annealing.  On similar inputs designed to suppress local degeneracy, performance of a quantum annealing processor on hard instances improves by orders of magnitude at the 512-qubit scale, while classical performance remains relatively unchanged.  Simulations indicate that perturbative crossings are the primary factor contributing to these heavy tails, while sensitivity to Hamiltonian misspecification error plays a less significant role in this particular setting.
\end{abstract}

\maketitle

\section{Introduction}

The recent development of quantum annealing (QA) processors has spurred research into how to construct input instances---Ising spin instances---that might confirm or refute any purported computational advantage conferred by such hardware \cite{McGeoch2013,Roennow2014,Venturelli2015a,Boixo2015,Denchev2015,henfl,flsat,Katzgraber2014,Katzgraber2015,Martin2015}.  These recent works have proposed various important considerations such as error sensitivity, thermal hardness \cite{Martin2015}, ground state degeneracy, consistency of energy scale across a random ensemble \cite{henfl,flsat}, potential barrier shape \cite{Boixo2015,Denchev2015}, and the existence of classical phase transitions \cite{Katzgraber2014}.  In parallel to this line of research is the consideration of suitable performance metrics for exact and approximate optimization \cite{ttt,Katzgraber2015,Steiger2015}.

Several previous efforts to quantify the performance of D-Wave hardware relative to physically motivated software solvers have focused on the median case performance for a given input class \cite{Roennow2014,henfl}.  However, Ref.\ \cite{Steiger2015} cautions that the median case does not give a complete picture of a given solver's performance.  Simulated quantum annealing (SQA) \cite{Santoro2002,boixo2014evidence}, which has been advanced as an effective model for D-Wave hardware, shows dramatically varying performance over particular random ensembles of low precision problems: the total time required to solve all instances is dominated by a few instances at the tail of the distribution.  In contrast, simulated annealing (SA) does not show nearly as strong of a variation in performance when run on the very same instances.  The presence of so-called {\em heavy tails} in the distribution of the time to solution, as seen in both SQA simulations \cite{Steiger2015} and D-Wave hardware results \cite{boixo2014evidence,Roennow2014}, points to an intrinsic physical mechanism that could explain the difference in performance between classical-physics-based approaches such as SA and the quantum-physics-based approaches such as SQA and D-Wave hardware.

In this paper we examine the hypothesis that {\em local degeneracy}---the abundance of large isoenergetic clusters of spin states---plays a crucial role in the appearance of heavy tails.  We compare the performance of a D-Wave 2X quantum anealing processor (DW2X) \cite{ttt} with SA on random $\pm 1$ Ising spin glass instances with high local degeneracy and low local degeneracy, where local degeneracy is controlled via parity of qubit connectivity rather than selection of coupling strengths \cite{Katzgraber2015,Zhu2015}.  Our results indicate that heavy tails are a consequence of degeneracy in the final target Hamiltonian and a particular implementation of (S)QA wherein all qubits follow an identical annealing schedule.  Under these circumstances, small gap perturbative anticrossings appear in the system eigenspectrum late in the anneal, at which point tunneling is suppressed.  These circumstances arise when random low-precision Ising spin glass instances are imposed on the full DW2X qubit connectivity graph.  However, they do not arise when the connectivity of that graph is varied as described herein, and they become less common when more realistic \cite{Rieffel2015,Venturelli2015,Babbush2014,Perdomo-Ortiz2015a} or combinatorially interesting problem classes are subject to (S)QA \cite{Venturelli2015a,Zdeborova2008}.

\section{Quantum annealing processor}

\subsection{Hamiltonians in the Ising model}

Quantum annealing in the Ising model aims to find low-energy states in a system of $n$ interacting spins via evolution of the time-dependent Hamiltonian
\begin{equation}\label{eq:ham}
  \begin{split}
    \fH_{S}(s)&=\frac{\epsilon(s)}{2}\fH_P - \frac{\Delta(s)}{2}\sum_{i}\sigma_i^x \\
    \fH_P &= \sum_{i<j}J_{ij}\sigma_i^z\sigma_j^z + \sum_i h_i\sigma_i^z.
  \end{split}
\end{equation}
Here $s \in [0,1]$ is the annealing parameter indicating progress through the anneal; $\sigma_i^x$ and $\sigma_i^z$ are Pauli matrices acting on spin $i$; $\Delta(s)/2$ and $\epsilon(s)/2$ are the energy scales corresponding to the transverse and longitudinal fields, respectively, and satisfy $\Delta(0)\gg\epsilon(0)$ and $\Delta(1)\ll \epsilon(1)$; $\fH_P$ is the Ising problem Hamiltonian \cite{Johnson2011}. {An isolated qubit with a bias $h=1$ has an energy gap of $\Delta(s)+\epsilon(s)$ between the two eigenvalues of its Hamiltonian $\fH_{S}(s)$, where $\Delta(s)$ is the contribution due to tunneling and $\epsilon(s)$ is the contribution due to the Ising Hamiltonian.}



\subsection{Floppy qubits and degeneracy in $J=\pm 1$ Chimera-structured Ising instances\label{sec:floppy}}

\newcommand{\heff}{h_i^{{\rm eff}}(\vec s)}

\begin{figure}
\includegraphics[width=.7\linewidth]{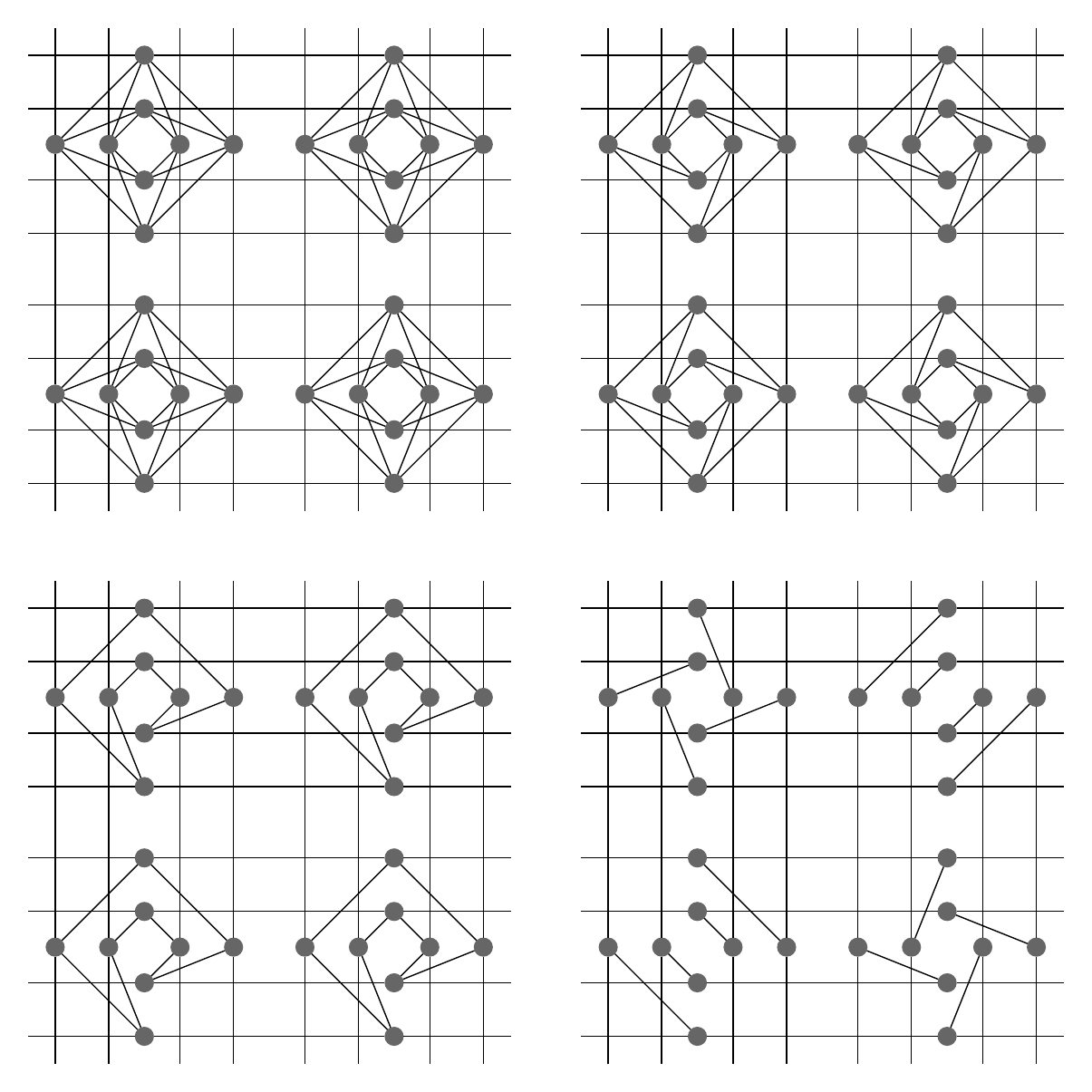}
\caption{Example subgraphs of a Chimera graph with varying degree.  Points indicate qubits and lines indicate couplers.  Top left is a section of a full Chimera graph, with maximum degee 6.  We can disable couplers in order to change the degree, and therefore floppiness, of the qubits.  Shown are reductions to target degree 5 (top right), 4 (bottom left) and 3 (bottom right).\label{fig:chimeradegrees}}
\end{figure}

Qubits in a D-Wave processor are connected as a {\em Chimera} graph \cite{bunyk2014architectural} (see Fig.\ \ref{fig:chimeradegrees}), in which most qubits are coupled to six others, provided that only a few qubits and couplers are inoperable.  For a classical $n$-qubit state $\vec s = \{s_1,\ldots,s_n\}\in\{-1,1\}^n$, each qubit $s_i$ experiences an {\em effective field}
\begin{equation}
\heff = h_i + \sum_{j> i}J_{ij}s_j.
\end{equation}
In the case $\heff=0$, we say that qubit $s_i$ is {\em floppy} in $\vec s$; equivalently, changing $s_i$ does not change the energy of the state.  Floppy qubits and the resulting degeneracy have a well-understood contribution to perturbative crossings \cite{Dickson2011,Dickson2013}, and they have been studied from the perspective of differentiating heuristics that operate over time-dependent quantum versus thermal annealing potential \cite{Boixo2013,Albash2015}.  In short, floppy qubits lead to wide, degenerate valleys in the state space that are favored in the time-dependent quantum potential early in the anneal.  We now illustrate the effect of floppy qubits on degeneracy in the random Ising instances often used to study D-Wave processors \cite{McGeoch2013,boixo2014evidence,Roennow2014,Steiger2015,ttt}.

\begin{figure}
\includegraphics[width=\linewidth]{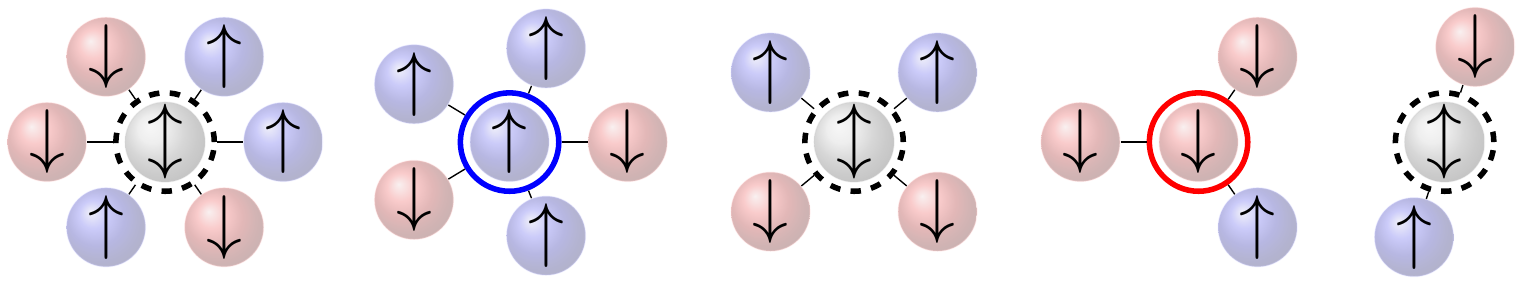}
\caption{A qubit $s_i$ is floppy in a state $\vec s$ if there is zero effective field on the qubit, i.e.\ $\heff=0$.  In this case changing $\sigma_i^z$ incurs no change in energy.  In the configurations shown, all couplings are ferromagnetic with value $-1$.  In $\pm 1$ problems, only qubits with even degree can be floppy (indicated with dashed outline).\label{fig:floppy}}
\end{figure}

Let $(h,J)$ be a random Ising spin glass instance with $h=\vec 0$ and with each operable coupler $J_{ij}$ taking a value in $\{-1,1\}$ uniformly at random.  A qubit $s_i$ coupled to $d$ others (i.e.\ of {\em degree $d$}), has $\Pr[\heff =0] = {d\choose {d/2}}/2^d$ when $d$ is even and $\Pr[\heff =0] =0$ when $d$ is odd (Fig.\ \ref{fig:floppy}).  Qubits of degree $2$, $4$, and $6$ have respective probabilities $50\%$, $38\%$, and $31\%$ of being floppy in a random spin configuration.  This probability approaches zero for large degree.  To put these probabilities in perspective, in a $512$-qubit Chimera-structured instance there are $384$ degree 6 qubits, of which we expect $120$ to be floppy in a random state.  This results in large isoenergetic clusters: due to the bipartite structure of Chimera, we expect every qubit in such a state to be in an isoenergetic hypercube of size $2^{60}$ \footnote{Given $120$ floppy qubits in a bipartite graph such as Chimera, there is a {\em stable set} $S$ of at least $60$ qubits with no couplings between them; the set of states reached by flipping any subset of $S$ is an isoenergetic hypercube of size $2^{|S|}$ in Hamming space.}.  Degeneracy will be less severe in local minima, where effective fields are not randomly distributed---for highly degenerate instances studied, roughly 5--10\% of qubits in low-lying excited states were floppy.

For positive integer $k$ we denote by $U_k$ the class of instances in which each coupler is set to a value in $\{\pm 1,\pm 2,\ldots, \pm k\}$ chosen uniformly at random.  Refs.\ \cite{Katzgraber2015,Zhu2015} avoided the local degeneracy seen in $U_k$ instances by drawing coupling strengths from {\em Sidon sets}; for example $S_{28}$ instances draw couplings from $\{\pm 28,\pm 19,\pm 13, \pm 8\}$.  Although this construction is effective in reducing local degeneracy, it brings a practical difficulty: Input to D-Wave processors must be rescaled so that entries of $h$ and $J$ have absolute value at most $1$, so $S_{28}$ instances must be rescaled by a factor of $1/28$.  Compared with $U_1$ instances, $S_{28}$ instances are therefore subject to a 28-fold inflation of relative noise, control error, and thermal effects relative to the final (classical) gap when run on DW2.  This effect increases for random spin glasses as the classical gap shrinks \cite{Zhu2015,flsat}, and decreases as excitations from ground state are accepted as viable solutions \cite{Katzgraber2015}.  This paper demonstrates that even a 5-fold reduction of energy scale significantly degrades DW2X performance on low-degeneracy $U_1$ instances (see Fig.\ \ref{fig:tails1}).

In this work we employed an alternative construction that suppresses or enhances local degeneracy without shrinking the classical gap through normalization.  First we simply disabled, by setting to zero, a subset of couplers that heuristically minimizes or maximizes the number of qubits with even degree.  We then imposed a $U_1$ instance on the remaining couplers.  More specifically, we chose a target degree $d\in \{3,4,5,6\}$ and attempted to make as many qubits as possible with this degree, while retaining the underlying gridlike structure of the graph by leaving inter-cell couplers mostly intact \footnote{We avoid degree 2 qubits because in $U_1$ problems they can be reduced, as {\em subdivisions} of smaller problems \cite{Diestel2012}, and therefore only add complexity in the form of degenerate subspaces.}.  We denote by $U_k^d$ the class of instances in which a $U_k$ instance is imposed on a subgraph with target degree $d$; note that $U_k = U_k^6$.  Ideal subgraph examples are shown in Fig.\ \ref{fig:chimeradegrees}.  The study of DW2X and SA performance on random instances with controlled degree (and consequently floppiness) illuminates the role of local degeneracy in the emergence of heavy tails.

\section{Results}

The testbed consisted of random $U_1$ and $U_4$ Ising instances on Chimera subgrids of varying size, from $3\times 3$ unit cells ($\fC_3$) to $10\times 10$ unit cells ($\fC_{10}$).  These problems contain between $72$ and $765$ operable qubits.  For each problem size, range $k$, and target degree $d$, we constructed $1000$ $U_k^d$ instances, each on a newly generated subgraph with target degree $d$ \footnote{Instances were generated on distinct random graphs to minimize the influence of structural anomalies that might arise in particular graphs.}.  We used the same processor and qubit configuration as Ref.\ \cite{ttt}.  Further details are given in the appendix.

\subsection{The smoking gun: Heavy tails for $U_1^4$ instances, but not $U_1^3$ instances}

\begin{figure}
\setlength{\figureheight}{3.5cm}%
\setlength{\figurewidth}{3.3cm}%
\includegraphics{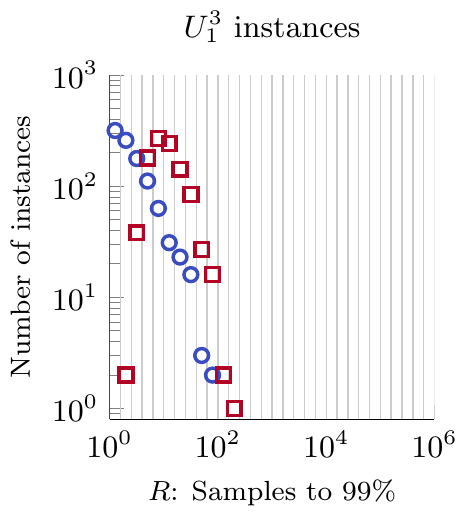}%
\includegraphics{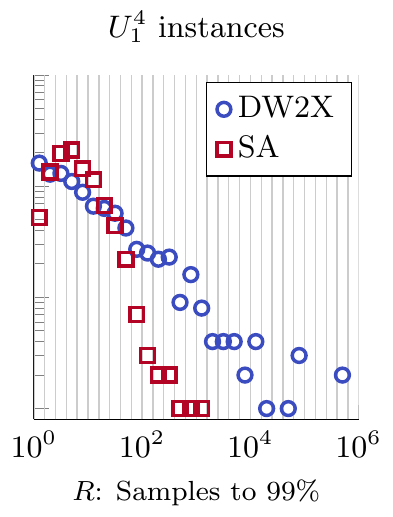}\\\vspace{.2cm}
\setlength{\figureheight}{3cm}%
\setlength{\figurewidth}{1.4cm}%
\hspace{-.2cm}%
\includegraphics{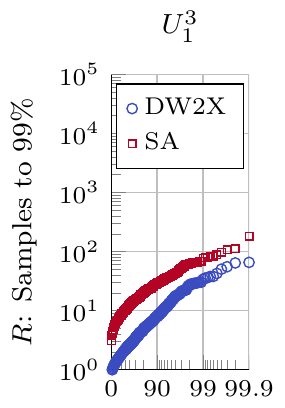}%
\includegraphics{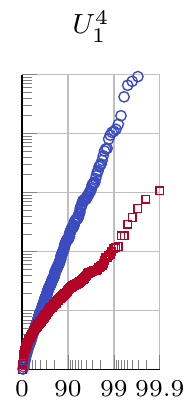}%
\includegraphics{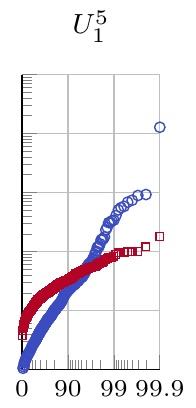}%
\includegraphics{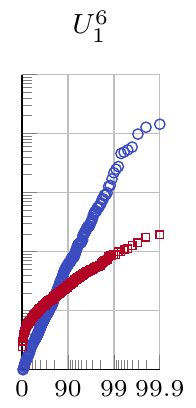}\\

Hardness percentile

\caption{Distribution of hardness for odd parity $U^3_1$ and even parity $U^4_1$ instances on 127 qubits ($\fC_4$).  (Top) Results for $U^3_1$ instances show comparable tails at large $R$ (low success probability) for DW2X and SA.  For $U^4_1$ instances, the DW2X results show the same characteristic heavy tail seen for SQA in Ref. \cite{Steiger2015}, whereas SA ($2^{10}$ sweeps, chosen to give comparable results to DW2X) shows no such tail.  (Bottom) We visualize these data with $R$ on the $y$-axis.  DW2X performance is highly dependent on degree.  For $U_1^4$ instances, DW2X and SA cross near the median, while for $U_1^3$ instances, DW2X dominates throughout the distribution.\label{fig:smokinggun}}
\end{figure}

Ref.\ \cite{Steiger2015} ran SQA and SA on the same 200-qubit ($\fC_5$) $U_1^6$ instances, each containing 120 degree 6 (and therefore potentially floppy) qubits, and investigated the ground state probability $p$ of each solver on a given instance.  While they consider $\tau = 1/p$, we consider the number of repetitions $R$ needed to achieve 99\% probability of finding a ground state \cite{boixo2014evidence}: 
\begin{equation}\label{eq:R}
R = \log(1-0.01)/\log(1-p).
\end{equation}
The two solvers exhibit similar success probability in the median case, but for the hardest instances SA has success probability orders of magnitude larger than SQA \cite{Steiger2015}.  If these ``heavy tails'' appear for QA across broad selections of input classes, it would call into question the potential utility of the general QA implementation given in Eq.\ \ref{eq:ham}.  Fig.\ \ref{fig:smokinggun} shows that on our inputs, DW2X shows a heavy tail on $U_1^4$ instances but not on $U_1^3$ instances.
This difference is not observed in SA results, which indicates that degree and degeneracy play a different role in QA than in SA.

The $U_1^6$ instances here have many degree 5 vertices, which may explain why they show less of a heavy tail than $U_1^4$ instances.  As for why $U_1^5$ instances show more of a heavy tail than $U_1^3$ instances, the cause is unknown but we provide one possible explanation:  $U_1^5$ instances contain more active couplers inside each unit cell of the Chimera graph, which may lead to the development of strongly coupled domains of qubits in unit cells; increased density of internal coupling has the effect of suppressing the effective transverse field on these domains, reducing their tunneling rate in the vicinity of a late anticrossing.

\subsection{Performance scaling on bimodal spin glasses}

\begin{figure}
\setlength{\figurewidth}{3cm}%
\setlength{\figureheight}{3cm}%
\includegraphics{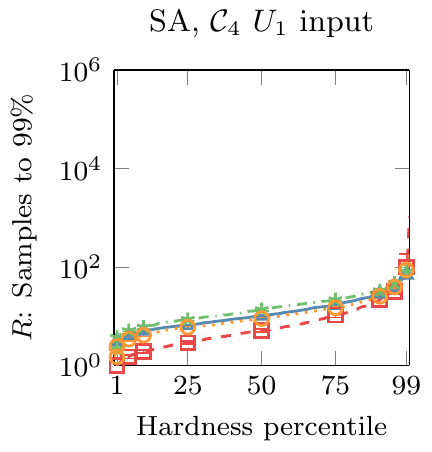}%
\includegraphics{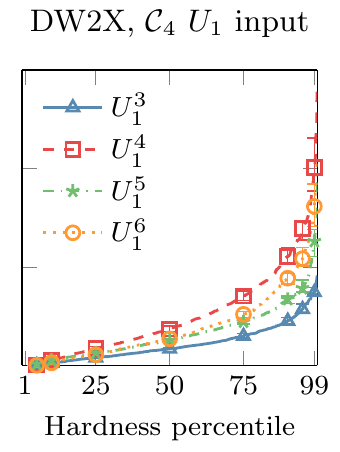}\\\vspace{.2cm}
\includegraphics{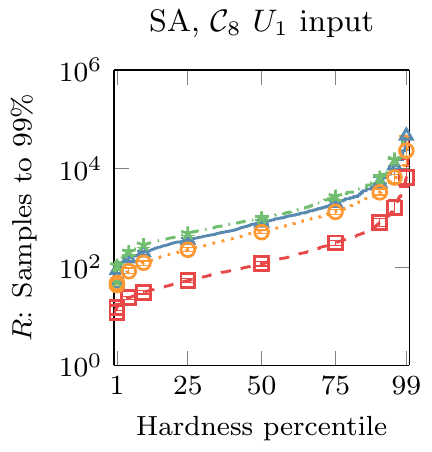}%
\includegraphics{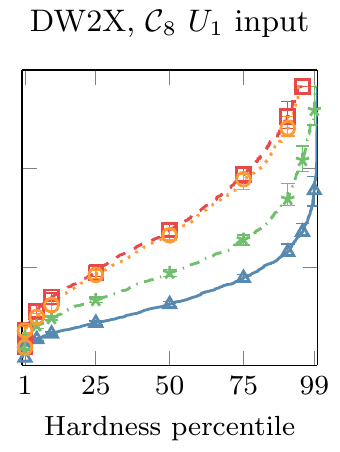}\\
\caption{(Top) Data from Fig.\ \ref{fig:smokinggun} rearranged with a linear $x$-axis.  Performance of DW2X relative to SA on hard instances is far better on $U_1^3$ and $U_1^5$ instances, where qubits with zero effective field are rare, than on the $U_1^6$ instances studied in recent work \cite{Roennow2014,Steiger2015,Katzgraber2015}.  Data points and error bars represent 95\% confidence intervals from bootstrap samples.  (Bottom) Corresponding plots for $\fC_8$ instances, using 498 available qubits out of 512.\label{fig:c4}}
\end{figure}

Fig.\ \ref{fig:c4} shows performance of DW2X and SA at the $\fC_4$ and $\fC_8$ scales, which were the respective sizes of of D-Wave One and D-Wave Two processors \cite{boixo2014evidence,Roennow2014}.
Again, performance of DW2X relative to SA improves enormously in high percentiles for $U_1^3$ instances, where removal of couplers minimizes degeneracy, versus $U_1^4$ instances.  Fig.\ \ref{fig:scaling1} shows performance scaling on $U_1^3$, $U_1^4$, $U_1^5$, and $U_1^6$ instances.  These results suggest that several threads of research that probed for computational advantage in $U_1^6$ problems \cite{Katzgraber2015,Martin2015,Roennow2014} might be more successfully directed towards $U_1^3$ problems.  Neither SA nor DW2X performance responds monotonically to increasing degree.  Thus the difficulty of a $U_1^d$ testbed as seen by QA or SA is not trivially related to the number of couplings it contains---additional couplings do not necessarily make a problem more or less difficult.

\subsection{Mean-field spectra of $U_1$ instances}

Heavy tails in problem sets with high local degeneracy might be explained by perturbative crossings late in the anneal for hard instances \cite{Dickson2011,Dickson2013}, when a superposition of many excited classical states ceases to be an instantaneous quantum ground state \cite{Amin2008,Amin2009a,Altshuler2010,Young2010}.  This crossing results in a minimum eigengap late in the anneal, when the tunneling rate is small.  As a result the annealer tends to get stuck in a local minimum, leading to low success probability.

Here we illustrate the relationship between gap position and local degeneracy.  For $127$-qubit $U_1$ instances we cannot determine the gap position using exact or even approximate diagonalization, as the instances have too many low-energy states.  As an alternative we resort to the mean-field spin vector Hamiltonian \cite{Klauder1979,Smolin2013} derived in Ref.\ \cite{Albash2015b} as the semiclassical limit of the spin-coherent states path integral \cite{Shin2014,Muthukrishnan2015}.  Qubits are represented by rotors on the unit circle, while $\sigma_x$, $\sigma_y$, and $\sigma_z$ are replaced with $\sin\theta$, $0$, and $\cos\theta$ respectively:
\begin{equation}\label{eq:sssvham}
\begin{split}
\fH_{{\rm SV}}(s)=&  \frac{\epsilon(s)}{2}\Bigg(\sum_{i<j}J_{ij}\cos\theta_i\cos\theta_j + \sum_i h_i\cos\theta_i\Bigg)\\
& - \frac{\Delta(s)}{2}\sum_{i}\sin\theta_i
\end{split}
\end{equation}
Ref.\ \cite{Muthukrishnan2015} showed that for certain inputs, the minimum gap location given by $\fH_{{\rm SV}}(s)$ closely matches the location given by exact diagonalization.  Here we aim to associate ground and excited states of the classical Hamiltonian with global and local minima of the time-dependent semiclassical Hamiltonian $\fH_{{\rm SV}}(s)$, estimating the gap location as the last moment at which the global minimum of $\fH_{{\rm SV}}(s)$ corresponds to a classical excited state.

For any annealing parameter $s$ and rotor state $\vec\theta$, we can map $\vec\theta$ to a local minimum of $\fH_{{\rm SV}}(s)$, denoted $L_s(\vec\theta)$, using a form of gradient descent (details are given in the appendix).  If we take $\vec\theta$ as a rotor state in $\{0,\pi\}^n$ that naturally represents a classical state, then early in the anneal, $L_s(\vec\theta)$ be an instantaneous ground state in the direction of the transverse field.  Late in the anneal, the transverse field will have little effect, so $L_s(\vec\theta)$ will only be an instantaneous ground state of $\fH_{{\rm SV}}(s)$ if $\vec\theta$ represents a classical ground state.

Computing $L_s(\vec\theta)$ for a range of $s$ and for $\vec\theta$ representing many classical ground and excited states gives a heuristic idea of when instantaneous and final ground states coincide according to the mean-field Hamiltonian.  We define the {\em mean-field crossing time} $s^*_{{\rm SV}}$ as the maximum value of the annealing parameter $s$ for which no $\vec\theta$ with $L_s(\vec\theta)$ minimizing $\fH_{{\rm SV}}(s)$ corresponds to a classical ground state.  We use $s^*_{{\rm SV}}$ as a mean-field approximation of the location of the minimum gap of the quantum Hamiltonian $\fH_{S}$.  This approach assumes that the problem has a single small gap resulting from an avoided level crossing, for example between a large metastable valley and a valley containing a classical ground state.

\begin{figure}
\begin{flushright}
\setlength{\figurewidth}{6.9cm}%
\setlength{\figureheight}{1.8cm}%
\includegraphics{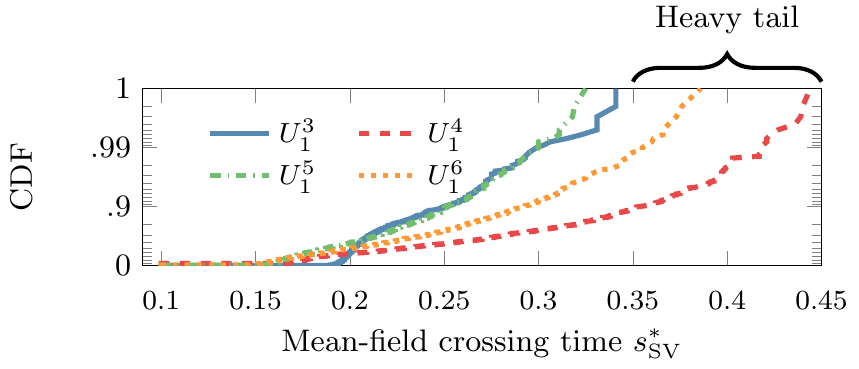}%
\\%
\setlength{\figurewidth}{6.9cm}%
\setlength{\figureheight}{4cm}%
\includegraphics{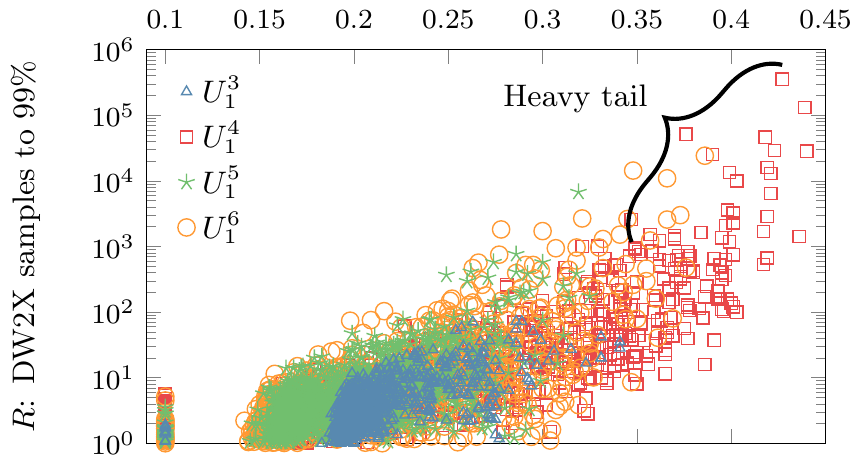}
\end{flushright}
\caption{Estimates of mean-field crossing time $s^*_{{\rm SV}}$ for 1000 127-qubit $\fC_4$ instances of $U_1^k$ for $k\in\{3,4,5,6\}$.  Instances for which no crossing was found are shown at $s=0.10$.  (Top) Empirical cumulative distribution function of $s^*_{{\rm SV}}$ shows the prominence of heavy tails for $U_1^4$ and $U_1^6$ instances.  (Bottom) DW2X success probability correlates well with $s^*_{{\rm SV}}$.  Hard instances have late crossing, suggesting the presence of perturbative crossings.  The ``heavy tail'' of hard instances, seen primarily for $U_1^4$ and $U_1^6$ instances, corresponds roughly to instances with $s^*_{{\rm SV}}>0.35$; transverse field is effectively off by $s=0.7$ (see Fig.\ \ref{fig:schedule}).\label{fig:svgaps}}
\end{figure}

Figure \ref{fig:svgaps} shows distributions of $s^*_{{\rm SV}}$ for instances of different degrees.  There is a strong correlation between a late crossing and low DW2X success probability.  Typically, a very late $s^*_{{\rm SV}}$ implies a crossing between a semiclassical superposition of many low-energy excited states and a relatively small number of ground states \cite{Dickson2013}.
Here $U_1^3$ and $U_1^5$ instances show very similar tail shapes, in contrast to DW2X output.  This may simply be due to a limitation of the semiclassical model, which does not reflect the tunneling dynamics in a quantum annealing processor.

\subsection{Tails and energy scale}

SA, DW2X, and other physically-motivated solvers are sensitive to multiplication of the problem Hamiltonian $\fH_P$ by a scaling factor $\alpha$.  Reducing $\alpha$ in SA is equivalent to increasing temperature, but in DW2X there are additional considerations of noise, error, and transverse field. 

\begin{figure}
  \setlength{\figurewidth}{1.4cm}%
  \setlength{\figureheight}{3cm}%
  \hspace{-.2cm}\includegraphics{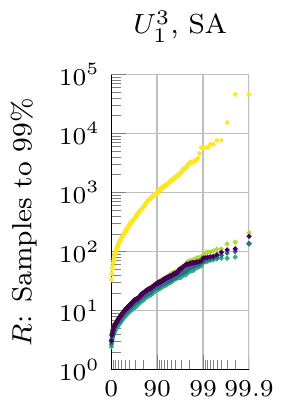}\includegraphics{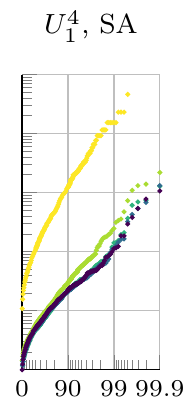}\includegraphics{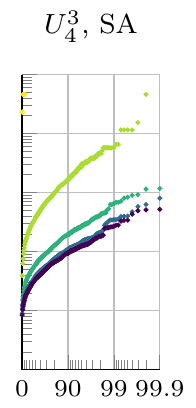}\includegraphics{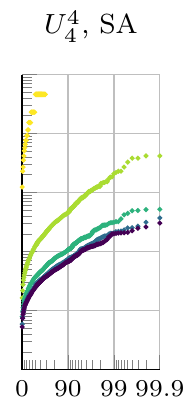}\\  
  Hardness percentile\vspace{2mm}\\
  \hspace{-.2cm}\includegraphics{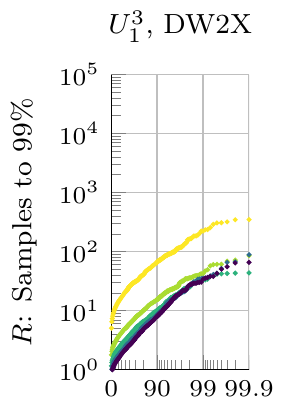}\includegraphics{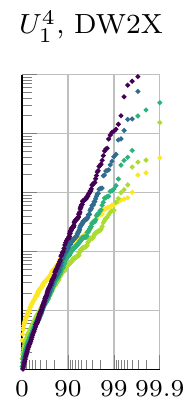}\includegraphics{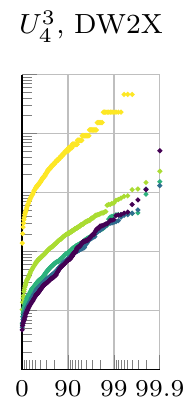}\includegraphics{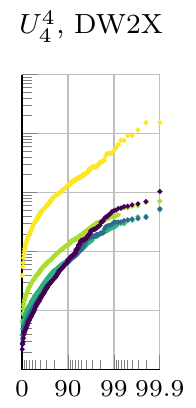}\\  
  Hardness percentile\vspace{3mm}\\

  Energy scale\\

  \includegraphics{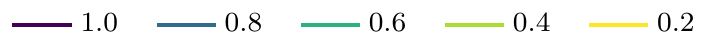}
  \caption{Tail shapes for SA (top) and DW2X (bottom) on $\fC_4$ 127-qubit $U_1^3$, $U_1^4$, $U_4^3$, and $U_4^4$ instances.  SA is run for $2^{10}$ sweeps per sample.  The problem Hamiltonian $\fH_P$ is multiplied by a prefactor ranging from $0.2$ to $1$.  Running at $0.2$ energy scale reduces the running time of DW2X by more than 100 times on the hardest $U_1^4$ problems, while having no positive effect on $U_1^3$ problems.  Instances in $U_4$ react similarly to changes in energy scale for both SA and DW2X.\label{fig:tails1}}
\end{figure}

 When multi-qubit tunneling is the dominant mechanism of solution for DW2X (likely not the case for $U_k^d$ instances \cite{Katzgraber2014,Shin2014,Albash2015b}), performance is best when $\alpha$ is maximized and relative temperature is minimized \cite{Boixo2015,Denchev2015}.  In contrast, Fig.\ \ref{fig:tails1} shows that heavy tails become less prevalent in DW2X results for $U_1^4$ input as we reduce $\alpha$.  This may be due to thermal excitation prior to diabatic Landau-Zener transition \cite{Zener1932} through a perturbative crossing, as hypothesized in Ref.\ \cite{Dickson2013}.  Reducing $\alpha$ degrades DW2X performance on easy $U_1^4$ and all $U_1^3$ instances, which agrees with the evidence in Fig.\ \ref{fig:svgaps} that these instances are not governed by late perturbative crossings.

Performance of DW2X on $U_4^3$ and $U_4^4$ instances is almost identical; in both regimes a qubit is floppy in a random state with probability only $7\%$, far lower than in $U_1^4$ instances.  Thus these instances are less degenerate, but are also run with couplings normalized by a factor of $\tfrac 14$, to within the range $[-1,1]$.  Even at $\alpha=0.2$, $U_1^4$ results shows a markedly more prominent tail as compared to $U_1^3$ results; the same is not true for $U_4^4$ and $U_4^3$ run at $\alpha=0.8$, where the classical gap is the same as $U_1$ instances at $\alpha=0.2$.

Simulated annealing, by contrast, sees relatively consistent performance at $\alpha=1$ across the input sets studied, and its performance degrades consistently as $\alpha$ is lowered relative to a fixed temperature schedule with $\beta_{{\rm final}}=5$ relative to $J\in[-1,1]$.

\subsection{Error sensitivity}

\begin{figure}
\setlength{\figurewidth}{2.8cm}%
\setlength{\figureheight}{3.5cm}
\begin{tabular}{rl}
  \includegraphics{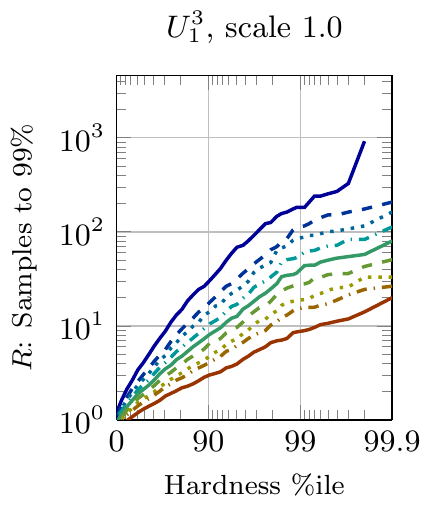}\hspace{.3cm} &
  \includegraphics{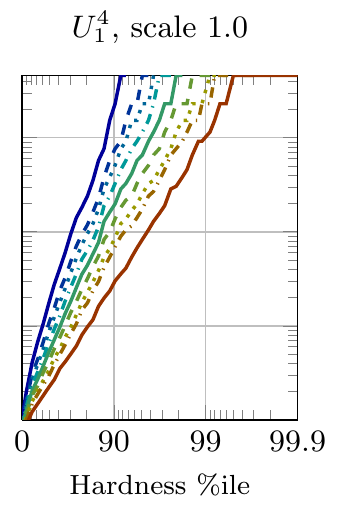}\\
  \includegraphics{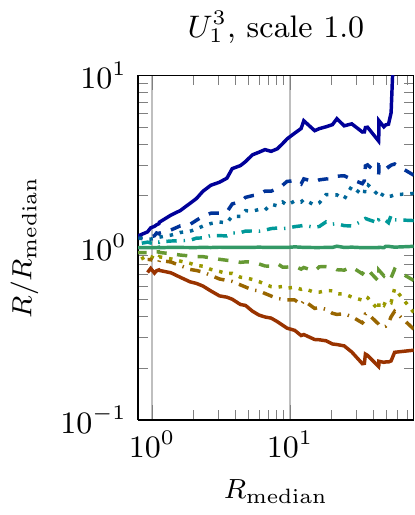}\hspace{.69cm} &
  \includegraphics{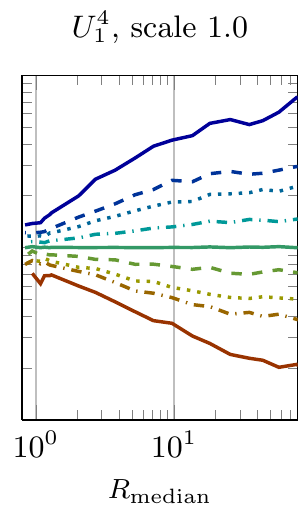}\\
\end{tabular}
\\
Percentile\\
  \includegraphics{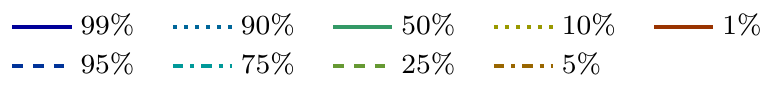}
  \caption{(Top) Varying DW2X performance on $U_1^3$ and $U_1^4$ $\fC_4$ instances; each instance is run using 10 spin reversal transformations, each of which is repeated 10 times for a total of 100 experiments per instance.  The experiment quantiles are then sorted by hardness as measured by $R$.  The middle line for example indicates the median of 100 experiments for each of 1000 instances.  $U_1^4$ results show a heavy tail at each experiment quantile, in contrast with $U_1^3$ results.  (Bottom) Results for each quantile are divided by median performance $R_{\textrm{median}}$, and plotted against $R_{\textrm{median}}$.  Bottom plots for $U_1^3$ and $U_1^4$ are restricted to the mutual domain.\label{fig:gauge1}.  Percentiles increase bottom to top.}
\end{figure}

Heavy tails in QA are most severe when energy scale is highest, i.e.\ when relative noise and error should be minimal.  The existence of heavy tails for SQA with no Hamiltonian misspecification \cite{Steiger2015} indicates that this phenomenon---and more generally, poor QA results on certain low-precision instances---cannot be solely attributed to control error.  To explore the contribution of classical control error to heavy tails, we ran our 127-qubit $U_1^3$ and $U_1^4$ instances 100 times each using 10 repetitions over random {\em spin reversal transformation} (otherwise known as {\em gauge transformations} \cite{Roennow2014,King2014}).  Fig.\ \ref{fig:gauge1} shows that heavy tails appear for $U_1^4$ across all quantiles (of 100 experiments), and do not appear at all for $U_1^3$ instances.

Clearly $R$ (Eq.\ \ref{eq:R}) varies more for hard problems than for easy problems.  More striking is how this spread of sorted performance quantiles is almost identical for $U_1^3$ and $U_1^4$ instances when normalized by the sorted median $R_{{\rm median}}$ (of 100 experiments) and plotted versus $R_{{\rm median}}$ rather than the percentile.  This indicates that heavy tails in DW2X results are not caused by error sensitivity, but rather that error sensitivity is closely related to hardness of the instance in terms of the quantum potential, rather than some classical measure of hardness derived from Boltzmann sampling, matrix condition number, or mixing time of a thermal process as studied in Ref.\ \cite{Martin2015}.  This perspective is further justified by heavy tails in SQA results \cite{Steiger2015} where the Hamiltonian is not prone to error, and by the fact that error sensitivity as measured by the spread of quantiles does not increase monotonically as energy scale decreases, increasing relative control error (see appendix).

\section{Discussion}

\subsection{Does local degeneracy challenge the viability of the QA platform?}

How important is local degeneracy in interesting problem classes?  Low-precision Chimera-structured problems have been used to benchmark D-Wave processors as a matter of simplicity and supposed resistance to control error.  The extreme local degeneracy in these instances arises from the fact that the Chimera qubit interaction graph is dominated by qubits of degree six, which is even.  When we disable couplers to suppress even-degree qubits, the tails associated with local degeneracy disappear.  This cannot be attributed to reduction of classical hardness: Figs.\ \ref{fig:c4} and \ref{fig:scaling1} show that for SA, $U_1^4$ instances are easier than $U_1^3$ and $U_1^5$ instances, while for DW2X the opposite is true.  This further indicates that the number of couplings has limited value as a measure of hardness in a random Ising problem.

Heavy tails are not present to the same extent in higher-precision instances, as we see in DW2X $U_4$ performance results.  However, we cannot dismiss the practical importance of low-precision problems, as there are hard low-precision combinatorial problems in the Ising model that are of more general interest than Chimera-structured $U_1$ instances.  Fully connected $\pm 1$ Sherrington-Kirkpatrick instances \cite{Sherrington1975,Venturelli2015a} are well understood and of interest beyond quantum annealing, but for a large instance each qubit is coupled to many others, and is therefore unlikely to be floppy regardless of the parity of its degree since ${d \choose \lceil d/2\rceil}/2^d$ scales as $d^{-1/2}$ (see Section \ref{sec:floppy}).  Any given small set of qubits is also unlikely to be floppy.  More recently there has been interest in {\em locked constraint satisfaction problems} \cite{Zdeborova2008,Zdeborova2008b,Zdeborova2011}, which are hard by design and have low-precision expressions in the Ising model \cite{Douglass2015}, for example 2-in-4-SAT and 3-in-6-SAT.  Locked constraint satisfaction problems explicitly avoid local degeneracy, and ground states are expected to be pairwise distant in Hamming space.  These constraint satisfaction problems tell us that hard, interesting problems in the Ising model do not need to have high precision or high degeneracy.

\subsection{Probing for speedup in highly degenerate instances}

Previous work has advocated searching for quantum speedup in hard $U_1^6$ problems \cite{Boixo2013,Roennow2014,Steiger2015,Martin2015}.  The relationship between degree and degeneracy provides two limitations of this choice, in addition to the fact that the classical potential of $U_1$ instances presents little challenge to simulated thermal solvers \cite{Katzgraber2014}.  First, $U_1^6$ problems have enormous local degeneracy, resulting in perturbative crossings late in the anneal for a subset---a heavy tail---of instances.  Those instances in the 99th percentile for QA time to solution will be exactly those instances that experience very late crossings, and therefore exhibit weakest performance of QA rather than the strongest (cf.\ \cite{Boixo2015,Denchev2015}).  Second, $U_1$ instances have been assumed to be robust to control error, but we have shown that robustness should be evaluated in terms of the quantum potential rather than the classical potential.  Instances experiencing late crossings will be sensitive to error regardless of precision required to represent each term in the classical Hamiltonian.  When probing for quantum speedup in the hardest of a set of random problems, degeneracy should be minimized in order to avoid disproportionate focus on instances susceptible to failure through late perturbative crossings.

While highly degenerate $U_1$ instances may have limited potential to exhibit quantum speedup, their simple structure and susceptibility to perturbative crossings may prove useful in the development of methods for avoiding perturbative crossings that might appear in more subtle ways.

\subsection{Mitigation of perturbative crossings}

Difficulties associated with QA and local degeneracy will likely be impossible to avoid altogether as we relax the definitions of {\em local} and {\em degeneracy}.  If high-precision instances do not have floppy qubits, they will at least sometimes have {\em nearly} floppy qubits, and it is unreasonable to expect that the time-dependent QA eigenspectrum will accurately reflect classical energy levels throughout the anneal.  In this case, can we deal with degeneracy and near-degeneracy?  Refs.\ \cite{Dickson2011} and \cite{Zhuang2014} propose two methods to mitigate perturbative crossings by manipulating the relative degeneracy of ground and excited states via addition of ancillary qubits.  These perturbative crossings, which arise in the presence of degenerate clusters of excited classical states \cite{Altshuler2010,Amin2009a}, might also be circumvented by algorithmic \cite{Dickson2012} or random \cite{Farhi2011} choice of the initial Hamiltonian and qubit annealing trajectories.

\section{Conclusions}

The heavy tails observed in SQA performance on $U_1^6$ instances \cite{Steiger2015} also appear in physical QA \cite{boixo2014evidence,Roennow2014}.  We have shown that among $U_1$ and $U_4$ instances, only those with high local degeneracy ($U_1^4$ and $U_1^6$) exhibit heavy tails in QA.  Moving from the $U_1^6$ instances normally studied in the literature to $U_1^3$ instances yields enormous improvement of QA performance relative to SA.

Mean-field simulations indicate that the heavy tails in highly degenerate instances are caused by Landau-Zener transitions through perturbative crossings late in the anneal.  These transitions appear in problems with large clusters of first excited states \cite{Amin2008,Amin2009a,Altshuler2010}, and may be circumvented (or enhanced \cite{Dickson2013,Boixo2013}) using structural \cite{Dickson2011} and algorithmic \cite{Dickson2012} modifications to the naive quantum annealing algorithm in which every qubit follows the same annealing trajectory.

\section{Acknowledgments}

The authors are grateful to Richard Harris for helpful advice regarding this manuscript, and to Damian Steiger for useful discussions about degeneracy and heavy tails.  We thank Quntao Zhuang and Helmut Katzgraber for discussions on an earlier version of this paper.

\bibliography{bibtex}

\begin{thebibliography}{50}%
\makeatletter
\providecommand \@ifxundefined [1]{%
 \@ifx{#1\undefined}
}%
\providecommand \@ifnum [1]{%
 \ifnum #1\expandafter \@firstoftwo
 \else \expandafter \@secondoftwo
 \fi
}%
\providecommand \@ifx [1]{%
 \ifx #1\expandafter \@firstoftwo
 \else \expandafter \@secondoftwo
 \fi
}%
\providecommand \natexlab [1]{#1}%
\providecommand \enquote  [1]{``#1''}%
\providecommand \bibnamefont  [1]{#1}%
\providecommand \bibfnamefont [1]{#1}%
\providecommand \citenamefont [1]{#1}%
\providecommand \href@noop [0]{\@secondoftwo}%
\providecommand \href [0]{\begingroup \@sanitize@url \@href}%
\providecommand \@href[1]{\@@startlink{#1}\@@href}%
\providecommand \@@href[1]{\endgroup#1\@@endlink}%
\providecommand \@sanitize@url [0]{\catcode `\\12\catcode `\$12\catcode
  `\&12\catcode `\#12\catcode `\^12\catcode `\_12\catcode `\%12\relax}%
\providecommand \@@startlink[1]{}%
\providecommand \@@endlink[0]{}%
\providecommand \url  [0]{\begingroup\@sanitize@url \@url }%
\providecommand \@url [1]{\endgroup\@href {#1}{\urlprefix }}%
\providecommand \urlprefix  [0]{URL }%
\providecommand \Eprint [0]{\href }%
\providecommand \doibase [0]{http://dx.doi.org/}%
\providecommand \selectlanguage [0]{\@gobble}%
\providecommand \bibinfo  [0]{\@secondoftwo}%
\providecommand \bibfield  [0]{\@secondoftwo}%
\providecommand \translation [1]{[#1]}%
\providecommand \BibitemOpen [0]{}%
\providecommand \bibitemStop [0]{}%
\providecommand \bibitemNoStop [0]{.\EOS\space}%
\providecommand \EOS [0]{\spacefactor3000\relax}%
\providecommand \BibitemShut  [1]{\csname bibitem#1\endcsname}%
\let\auto@bib@innerbib\@empty
\bibitem [{\citenamefont {McGeoch}\ and\ \citenamefont
  {Wang}(2013)}]{McGeoch2013}%
  \BibitemOpen
  \bibfield  {author} {\bibinfo {author} {\bibfnamefont {C.}~\bibnamefont
  {McGeoch}}\ and\ \bibinfo {author} {\bibfnamefont {C.}~\bibnamefont {Wang}},\
  }in\ \href@noop {} {\emph {\bibinfo {booktitle} {Proceedings of the ACM
  International Conference on Computing Frontiers}}}\ (\bibinfo {organization}
  {ACM},\ \bibinfo {year} {2013})\ p.~\bibinfo {pages} {23}\BibitemShut
  {NoStop}%
\bibitem [{\citenamefont {R{{\o}}nnow}\ \emph {et~al.}(2014)\citenamefont
  {R{{\o}}nnow}, \citenamefont {Wang}, \citenamefont {Job}, \citenamefont
  {Boixo}, \citenamefont {Isakov}, \citenamefont {Wecker}, \citenamefont
  {Martinis}, \citenamefont {Lidar},\ and\ \citenamefont
  {Troyer}}]{Roennow2014}%
  \BibitemOpen
  \bibfield  {author} {\bibinfo {author} {\bibfnamefont {T.}~\bibnamefont
  {R{{\o}}nnow}}, \bibinfo {author} {\bibfnamefont {Z.}~\bibnamefont {Wang}},
  \bibinfo {author} {\bibfnamefont {J.}~\bibnamefont {Job}}, \bibinfo {author}
  {\bibfnamefont {S.}~\bibnamefont {Boixo}}, \bibinfo {author} {\bibfnamefont
  {S.}~\bibnamefont {Isakov}}, \bibinfo {author} {\bibfnamefont
  {D.}~\bibnamefont {Wecker}}, \bibinfo {author} {\bibfnamefont
  {J.}~\bibnamefont {Martinis}}, \bibinfo {author} {\bibfnamefont
  {D.}~\bibnamefont {Lidar}}, \ and\ \bibinfo {author} {\bibfnamefont
  {M.}~\bibnamefont {Troyer}},\ }\href@noop {} {\bibfield  {journal} {\bibinfo
  {journal} {Science}\ }\textbf {\bibinfo {volume} {345}},\ \bibinfo {pages}
  {420} (\bibinfo {year} {2014})},\ \Eprint
  {http://arxiv.org/abs/http://www.sciencemag.org/content/345/6195/420.full.pdf}
  {http://www.sciencemag.org/content/345/6195/420.full.pdf} \BibitemShut
  {NoStop}%
\bibitem [{\citenamefont {Venturelli}\ \emph
  {et~al.}(2015{\natexlab{a}})\citenamefont {Venturelli}, \citenamefont
  {Mandr{\`a}}, \citenamefont {Knysh}, \citenamefont {O’Gorman},
  \citenamefont {Biswas},\ and\ \citenamefont {Smelyanskiy}}]{Venturelli2015a}%
  \BibitemOpen
  \bibfield  {author} {\bibinfo {author} {\bibfnamefont {D.}~\bibnamefont
  {Venturelli}}, \bibinfo {author} {\bibfnamefont {S.}~\bibnamefont
  {Mandr{\`a}}}, \bibinfo {author} {\bibfnamefont {S.}~\bibnamefont {Knysh}},
  \bibinfo {author} {\bibfnamefont {B.}~\bibnamefont {O’Gorman}}, \bibinfo
  {author} {\bibfnamefont {R.}~\bibnamefont {Biswas}}, \ and\ \bibinfo {author}
  {\bibfnamefont {V.}~\bibnamefont {Smelyanskiy}},\ }\href@noop {} {\bibfield
  {journal} {\bibinfo  {journal} {Phys. Rev. X}\ }\textbf {\bibinfo {volume}
  {5}},\ \bibinfo {pages} {031040} (\bibinfo {year}
  {2015}{\natexlab{a}})}\BibitemShut {NoStop}%
\bibitem [{\citenamefont {Boixo}\ \emph {et~al.}(2015)\citenamefont {Boixo},
  \citenamefont {Smelyanskiy}, \citenamefont {Shabani}, \citenamefont {Isakov},
  \citenamefont {Dykman}, \citenamefont {Denchev}, \citenamefont {Amin},
  \citenamefont {Smirnov}, \citenamefont {Mohseni},\ and\ \citenamefont
  {Neven}}]{Boixo2015}%
  \BibitemOpen
  \bibfield  {author} {\bibinfo {author} {\bibfnamefont {S.}~\bibnamefont
  {Boixo}}, \bibinfo {author} {\bibfnamefont {V.~N.}\ \bibnamefont
  {Smelyanskiy}}, \bibinfo {author} {\bibfnamefont {A.}~\bibnamefont
  {Shabani}}, \bibinfo {author} {\bibfnamefont {S.~V.}\ \bibnamefont {Isakov}},
  \bibinfo {author} {\bibfnamefont {M.}~\bibnamefont {Dykman}}, \bibinfo
  {author} {\bibfnamefont {V.~S.}\ \bibnamefont {Denchev}}, \bibinfo {author}
  {\bibfnamefont {M.}~\bibnamefont {Amin}}, \bibinfo {author} {\bibfnamefont
  {A.}~\bibnamefont {Smirnov}}, \bibinfo {author} {\bibfnamefont
  {M.}~\bibnamefont {Mohseni}}, \ and\ \bibinfo {author} {\bibfnamefont
  {H.}~\bibnamefont {Neven}},\ }\href@noop {} {\bibfield  {journal} {\bibinfo
  {journal} {arXiv preprint arXiv:1502.05754}\ } (\bibinfo {year}
  {2015})}\BibitemShut {NoStop}%
\bibitem [{\citenamefont {Denchev}\ \emph {et~al.}(2015)\citenamefont
  {Denchev}, \citenamefont {Boixo}, \citenamefont {Isakov}, \citenamefont
  {Ding}, \citenamefont {Babbush}, \citenamefont {Smelyanskiy}, \citenamefont
  {Martinis},\ and\ \citenamefont {Neven}}]{Denchev2015}%
  \BibitemOpen
  \bibfield  {author} {\bibinfo {author} {\bibfnamefont {V.~S.}\ \bibnamefont
  {Denchev}}, \bibinfo {author} {\bibfnamefont {S.}~\bibnamefont {Boixo}},
  \bibinfo {author} {\bibfnamefont {S.~V.}\ \bibnamefont {Isakov}}, \bibinfo
  {author} {\bibfnamefont {N.}~\bibnamefont {Ding}}, \bibinfo {author}
  {\bibfnamefont {R.}~\bibnamefont {Babbush}}, \bibinfo {author} {\bibfnamefont
  {V.}~\bibnamefont {Smelyanskiy}}, \bibinfo {author} {\bibfnamefont
  {J.}~\bibnamefont {Martinis}}, \ and\ \bibinfo {author} {\bibfnamefont
  {H.}~\bibnamefont {Neven}},\ }\href@noop {} {\bibfield  {journal} {\bibinfo
  {journal} {arXiv preprint arXiv:1512.02206}\ } (\bibinfo {year}
  {2015})}\BibitemShut {NoStop}%
\bibitem [{\citenamefont {Hen}\ \emph {et~al.}(2015)\citenamefont {Hen},
  \citenamefont {Albash}, \citenamefont {Job}, \citenamefont {R{\o}nnow},
  \citenamefont {Troyer},\ and\ \citenamefont {Lidar}}]{henfl}%
  \BibitemOpen
  \bibfield  {author} {\bibinfo {author} {\bibfnamefont {I.}~\bibnamefont
  {Hen}}, \bibinfo {author} {\bibfnamefont {T.}~\bibnamefont {Albash}},
  \bibinfo {author} {\bibfnamefont {J.}~\bibnamefont {Job}}, \bibinfo {author}
  {\bibfnamefont {T.~F.}\ \bibnamefont {R{\o}nnow}}, \bibinfo {author}
  {\bibfnamefont {M.}~\bibnamefont {Troyer}}, \ and\ \bibinfo {author}
  {\bibfnamefont {D.}~\bibnamefont {Lidar}},\ }\href@noop {} {\enquote
  {\bibinfo {title} {Probing for quantum speedup in spin glass problems with
  planted solutions},}\ } (\bibinfo {year} {2015}),\ \bibinfo {note} {arXiv
  preprint arXiv:1502.01663v2}\BibitemShut {NoStop}%
\bibitem [{\citenamefont {King}\ \emph
  {et~al.}(2015{\natexlab{a}})\citenamefont {King}, \citenamefont {Lanting},\
  and\ \citenamefont {Harris}}]{flsat}%
  \BibitemOpen
  \bibfield  {author} {\bibinfo {author} {\bibfnamefont {A.~D.}\ \bibnamefont
  {King}}, \bibinfo {author} {\bibfnamefont {T.}~\bibnamefont {Lanting}}, \
  and\ \bibinfo {author} {\bibfnamefont {R.}~\bibnamefont {Harris}},\
  }\href@noop {} {\enquote {\bibinfo {title} {Performance of a quantum annealer
  on range-limited constraint satisfaction problems},}\ } (\bibinfo {year}
  {2015}{\natexlab{a}}),\ \bibinfo {note} {arXiv preprint arXiv:1502.02098v2},\
  \Eprint {http://arxiv.org/abs/1502.02098v2} {arXiv:1502.02098v2} \BibitemShut
  {NoStop}%
\bibitem [{\citenamefont {Katzgraber}\ \emph {et~al.}(2014)\citenamefont
  {Katzgraber}, \citenamefont {Hamze},\ and\ \citenamefont
  {Andrist}}]{Katzgraber2014}%
  \BibitemOpen
  \bibfield  {author} {\bibinfo {author} {\bibfnamefont {H.~G.}\ \bibnamefont
  {Katzgraber}}, \bibinfo {author} {\bibfnamefont {F.}~\bibnamefont {Hamze}}, \
  and\ \bibinfo {author} {\bibfnamefont {R.~S.}\ \bibnamefont {Andrist}},\
  }\href@noop {} {\bibfield  {journal} {\bibinfo  {journal} {Phys. Rev. X}\
  }\textbf {\bibinfo {volume} {4}},\ \bibinfo {pages} {021008} (\bibinfo {year}
  {2014})}\BibitemShut {NoStop}%
\bibitem [{\citenamefont {Katzgraber}\ \emph {et~al.}(2015)\citenamefont
  {Katzgraber}, \citenamefont {Hamze}, \citenamefont {Zhu}, \citenamefont
  {Ochoa},\ and\ \citenamefont {Munoz-Bauza}}]{Katzgraber2015}%
  \BibitemOpen
  \bibfield  {author} {\bibinfo {author} {\bibfnamefont {H.~G.}\ \bibnamefont
  {Katzgraber}}, \bibinfo {author} {\bibfnamefont {F.}~\bibnamefont {Hamze}},
  \bibinfo {author} {\bibfnamefont {Z.}~\bibnamefont {Zhu}}, \bibinfo {author}
  {\bibfnamefont {A.~J.}\ \bibnamefont {Ochoa}}, \ and\ \bibinfo {author}
  {\bibfnamefont {H.}~\bibnamefont {Munoz-Bauza}},\ }\href {\doibase
  10.1103/PhysRevX.5.031026} {\bibfield  {journal} {\bibinfo  {journal} {Phys.
  Rev. X}\ }\textbf {\bibinfo {volume} {5}},\ \bibinfo {pages} {031026}
  (\bibinfo {year} {2015})}\BibitemShut {NoStop}%
\bibitem [{\citenamefont {Martin-Mayor}\ and\ \citenamefont
  {Hen}(2015)}]{Martin2015}%
  \BibitemOpen
  \bibfield  {author} {\bibinfo {author} {\bibfnamefont {V.}~\bibnamefont
  {Martin-Mayor}}\ and\ \bibinfo {author} {\bibfnamefont {I.}~\bibnamefont
  {Hen}},\ }\href@noop {} {\bibfield  {journal} {\bibinfo  {journal} {Nat. Sci.
  Rep.}\ }\textbf {\bibinfo {volume} {5}} (\bibinfo {year} {2015})}\BibitemShut
  {NoStop}%
\bibitem [{\citenamefont {King}\ \emph
  {et~al.}(2015{\natexlab{b}})\citenamefont {King}, \citenamefont {Yarkoni},
  \citenamefont {Nevisi}, \citenamefont {Hilton},\ and\ \citenamefont
  {McGeoch}}]{ttt}%
  \BibitemOpen
  \bibfield  {author} {\bibinfo {author} {\bibfnamefont {J.}~\bibnamefont
  {King}}, \bibinfo {author} {\bibfnamefont {S.}~\bibnamefont {Yarkoni}},
  \bibinfo {author} {\bibfnamefont {M.~M.}\ \bibnamefont {Nevisi}}, \bibinfo
  {author} {\bibfnamefont {J.~P.}\ \bibnamefont {Hilton}}, \ and\ \bibinfo
  {author} {\bibfnamefont {C.~C.}\ \bibnamefont {McGeoch}},\ }\href@noop {}
  {\enquote {\bibinfo {title} {{Benchmarking a quantum annealing processor with
  the time-to-target metric}},}\ } (\bibinfo {year} {2015}{\natexlab{b}}),\
  \Eprint {http://arxiv.org/abs/1508.05087v1} {arXiv:1508.05087v1} \BibitemShut
  {NoStop}%
\bibitem [{\citenamefont {Steiger}\ \emph {et~al.}(2015)\citenamefont
  {Steiger}, \citenamefont {R\o{}nnow},\ and\ \citenamefont
  {Troyer}}]{Steiger2015}%
  \BibitemOpen
  \bibfield  {author} {\bibinfo {author} {\bibfnamefont {D.~S.}\ \bibnamefont
  {Steiger}}, \bibinfo {author} {\bibfnamefont {T.~F.}\ \bibnamefont
  {R\o{}nnow}}, \ and\ \bibinfo {author} {\bibfnamefont {M.}~\bibnamefont
  {Troyer}},\ }\href {\doibase 10.1103/PhysRevLett.115.230501} {\bibfield
  {journal} {\bibinfo  {journal} {Phys. Rev. Lett.}\ }\textbf {\bibinfo
  {volume} {115}},\ \bibinfo {pages} {230501} (\bibinfo {year}
  {2015})}\BibitemShut {NoStop}%
\bibitem [{\citenamefont {Santoro}\ \emph {et~al.}(2002)\citenamefont
  {Santoro}, \citenamefont {Marto{\v{n}}{\'a}k}, \citenamefont {Tosatti},\ and\
  \citenamefont {Car}}]{Santoro2002}%
  \BibitemOpen
  \bibfield  {author} {\bibinfo {author} {\bibfnamefont {G.~E.}\ \bibnamefont
  {Santoro}}, \bibinfo {author} {\bibfnamefont {R.}~\bibnamefont
  {Marto{\v{n}}{\'a}k}}, \bibinfo {author} {\bibfnamefont {E.}~\bibnamefont
  {Tosatti}}, \ and\ \bibinfo {author} {\bibfnamefont {R.}~\bibnamefont
  {Car}},\ }\href@noop {} {\bibfield  {journal} {\bibinfo  {journal} {Science}\
  }\textbf {\bibinfo {volume} {295}},\ \bibinfo {pages} {2427} (\bibinfo {year}
  {2002})}\BibitemShut {NoStop}%
\bibitem [{\citenamefont {Boixo}\ \emph {et~al.}(2014)\citenamefont {Boixo},
  \citenamefont {R{{\o}}nnow}, \citenamefont {Isakov}, \citenamefont {Wang},
  \citenamefont {Wecker}, \citenamefont {Lidar}, \citenamefont {Martinis},\
  and\ \citenamefont {Troyer}}]{boixo2014evidence}%
  \BibitemOpen
  \bibfield  {author} {\bibinfo {author} {\bibfnamefont {S.}~\bibnamefont
  {Boixo}}, \bibinfo {author} {\bibfnamefont {T.}~\bibnamefont {R{{\o}}nnow}},
  \bibinfo {author} {\bibfnamefont {S.}~\bibnamefont {Isakov}}, \bibinfo
  {author} {\bibfnamefont {Z.}~\bibnamefont {Wang}}, \bibinfo {author}
  {\bibfnamefont {D.}~\bibnamefont {Wecker}}, \bibinfo {author} {\bibfnamefont
  {D.}~\bibnamefont {Lidar}}, \bibinfo {author} {\bibfnamefont
  {J.}~\bibnamefont {Martinis}}, \ and\ \bibinfo {author} {\bibfnamefont
  {M.}~\bibnamefont {Troyer}},\ }\href@noop {} {\bibfield  {journal} {\bibinfo
  {journal} {Nat. Phys.}\ }\textbf {\bibinfo {volume} {10}},\ \bibinfo {pages}
  {218} (\bibinfo {year} {2014})}\BibitemShut {NoStop}%
\bibitem [{\citenamefont {Zhu}\ \emph {et~al.}(2015)\citenamefont {Zhu},
  \citenamefont {Ochoa}, \citenamefont {Schnabel}, \citenamefont {Hamze},\ and\
  \citenamefont {Katzgraber}}]{Zhu2015}%
  \BibitemOpen
  \bibfield  {author} {\bibinfo {author} {\bibfnamefont {Z.}~\bibnamefont
  {Zhu}}, \bibinfo {author} {\bibfnamefont {A.~J.}\ \bibnamefont {Ochoa}},
  \bibinfo {author} {\bibfnamefont {S.}~\bibnamefont {Schnabel}}, \bibinfo
  {author} {\bibfnamefont {F.}~\bibnamefont {Hamze}}, \ and\ \bibinfo {author}
  {\bibfnamefont {H.~G.}\ \bibnamefont {Katzgraber}},\ }\href
  {http://arxiv.org/abs/1505.02278} {\enquote {\bibinfo {title} {{Best-case
  performance of quantum annealers on native spin-glass benchmarks: How chaos
  can affect success probabilities}},}\ } (\bibinfo {year} {2015}),\ \Eprint
  {http://arxiv.org/abs/1505.02278v1} {arXiv:1505.02278v1} \BibitemShut
  {NoStop}%
\bibitem [{\citenamefont {Rieffel}\ \emph {et~al.}(2015)\citenamefont
  {Rieffel}, \citenamefont {Venturelli}, \citenamefont {O'Gorman},
  \citenamefont {Do}, \citenamefont {Prystay},\ and\ \citenamefont
  {Smelyanskiy}}]{Rieffel2015}%
  \BibitemOpen
  \bibfield  {author} {\bibinfo {author} {\bibfnamefont {E.~G.}\ \bibnamefont
  {Rieffel}}, \bibinfo {author} {\bibfnamefont {D.}~\bibnamefont {Venturelli}},
  \bibinfo {author} {\bibfnamefont {B.}~\bibnamefont {O'Gorman}}, \bibinfo
  {author} {\bibfnamefont {M.~B.}\ \bibnamefont {Do}}, \bibinfo {author}
  {\bibfnamefont {E.~M.}\ \bibnamefont {Prystay}}, \ and\ \bibinfo {author}
  {\bibfnamefont {V.~N.}\ \bibnamefont {Smelyanskiy}},\ }\href@noop {}
  {\bibfield  {journal} {\bibinfo  {journal} {Quantum Information Processing}\
  }\textbf {\bibinfo {volume} {14}},\ \bibinfo {pages} {1} (\bibinfo {year}
  {2015})}\BibitemShut {NoStop}%
\bibitem [{\citenamefont {Venturelli}\ \emph
  {et~al.}(2015{\natexlab{b}})\citenamefont {Venturelli}, \citenamefont
  {Marchand},\ and\ \citenamefont {Rojo}}]{Venturelli2015}%
  \BibitemOpen
  \bibfield  {author} {\bibinfo {author} {\bibfnamefont {D.}~\bibnamefont
  {Venturelli}}, \bibinfo {author} {\bibfnamefont {D.~J.}\ \bibnamefont
  {Marchand}}, \ and\ \bibinfo {author} {\bibfnamefont {G.}~\bibnamefont
  {Rojo}},\ }\href {http://arxiv.org/abs/1506.08479} {\enquote {\bibinfo
  {title} {Quantum annealing implementation of job-shop scheduling},}\ }
  (\bibinfo {year} {2015}{\natexlab{b}}),\ \Eprint
  {http://arxiv.org/abs/1506.08479v1} {arXiv:1506.08479v1} \BibitemShut
  {NoStop}%
\bibitem [{\citenamefont {Babbush}\ \emph {et~al.}(2014)\citenamefont
  {Babbush}, \citenamefont {Perdomo-Ortiz}, \citenamefont {O'Gorman},
  \citenamefont {Macready},\ and\ \citenamefont {Aspuru-Guzik}}]{Babbush2014}%
  \BibitemOpen
  \bibfield  {author} {\bibinfo {author} {\bibfnamefont {R.}~\bibnamefont
  {Babbush}}, \bibinfo {author} {\bibfnamefont {A.}~\bibnamefont
  {Perdomo-Ortiz}}, \bibinfo {author} {\bibfnamefont {B.}~\bibnamefont
  {O'Gorman}}, \bibinfo {author} {\bibfnamefont {W.}~\bibnamefont {Macready}},
  \ and\ \bibinfo {author} {\bibfnamefont {A.}~\bibnamefont {Aspuru-Guzik}},\
  }\enquote {\bibinfo {title} {Construction of energy functions for lattice
  heteropolymer models: Efficient encodings for constraint satisfaction
  programming and quantum annealing},}\ in\ \href {\doibase
  10.1002/9781118755815.ch05} {\emph {\bibinfo {booktitle} {Advances in
  Chemical Physics: Volume 155}}}\ (\bibinfo  {publisher} {John Wiley \& Sons,
  Inc.},\ \bibinfo {year} {2014})\ pp.\ \bibinfo {pages} {201--244}\BibitemShut
  {NoStop}%
\bibitem [{\citenamefont {Perdomo-Ortiz}\ \emph {et~al.}(2015)\citenamefont
  {Perdomo-Ortiz}, \citenamefont {Fluegemann}, \citenamefont {Narasimhan},
  \citenamefont {Biswas},\ and\ \citenamefont
  {Smelyanskiy}}]{Perdomo-Ortiz2015a}%
  \BibitemOpen
  \bibfield  {author} {\bibinfo {author} {\bibfnamefont {A.}~\bibnamefont
  {Perdomo-Ortiz}}, \bibinfo {author} {\bibfnamefont {J.}~\bibnamefont
  {Fluegemann}}, \bibinfo {author} {\bibfnamefont {S.}~\bibnamefont
  {Narasimhan}}, \bibinfo {author} {\bibfnamefont {R.}~\bibnamefont {Biswas}},
  \ and\ \bibinfo {author} {\bibfnamefont {V.~N.}\ \bibnamefont
  {Smelyanskiy}},\ }\href@noop {} {\bibfield  {journal} {\bibinfo  {journal}
  {Eur. Phys. J. Spec. Top.}\ }\textbf {\bibinfo {volume} {224}},\ \bibinfo
  {pages} {131} (\bibinfo {year} {2015})}\BibitemShut {NoStop}%
\bibitem [{\citenamefont {Zdeborov{\'a}}\ and\ \citenamefont
  {M{\'e}zard}(2008{\natexlab{a}})}]{Zdeborova2008}%
  \BibitemOpen
  \bibfield  {author} {\bibinfo {author} {\bibfnamefont {L.}~\bibnamefont
  {Zdeborov{\'a}}}\ and\ \bibinfo {author} {\bibfnamefont {M.}~\bibnamefont
  {M{\'e}zard}},\ }\href@noop {} {\bibfield  {journal} {\bibinfo  {journal}
  {Phys. Rev. Lett.}\ }\textbf {\bibinfo {volume} {101}},\ \bibinfo {pages}
  {078702} (\bibinfo {year} {2008}{\natexlab{a}})}\BibitemShut {NoStop}%
\bibitem [{\citenamefont {Johnson}\ \emph {et~al.}(2011)\citenamefont
  {Johnson}, \citenamefont {Amin}, \citenamefont {Gildert}, \citenamefont
  {Lanting}, \citenamefont {Hamze}, \citenamefont {Dickson}, \citenamefont
  {Harris}, \citenamefont {Berkley}, \citenamefont {Johansson}, \citenamefont
  {Bunyk} \emph {et~al.}}]{Johnson2011}%
  \BibitemOpen
  \bibfield  {author} {\bibinfo {author} {\bibfnamefont {M.}~\bibnamefont
  {Johnson}}, \bibinfo {author} {\bibfnamefont {M.}~\bibnamefont {Amin}},
  \bibinfo {author} {\bibfnamefont {S.}~\bibnamefont {Gildert}}, \bibinfo
  {author} {\bibfnamefont {T.}~\bibnamefont {Lanting}}, \bibinfo {author}
  {\bibfnamefont {F.}~\bibnamefont {Hamze}}, \bibinfo {author} {\bibfnamefont
  {N.}~\bibnamefont {Dickson}}, \bibinfo {author} {\bibfnamefont
  {R.}~\bibnamefont {Harris}}, \bibinfo {author} {\bibfnamefont
  {A.}~\bibnamefont {Berkley}}, \bibinfo {author} {\bibfnamefont
  {J.}~\bibnamefont {Johansson}}, \bibinfo {author} {\bibfnamefont
  {P.}~\bibnamefont {Bunyk}},  \emph {et~al.},\ }\href@noop {} {\bibfield
  {journal} {\bibinfo  {journal} {Nature}\ }\textbf {\bibinfo {volume} {473}},\
  \bibinfo {pages} {194} (\bibinfo {year} {2011})}\BibitemShut {NoStop}%
\bibitem [{\citenamefont {Bunyk}\ \emph {et~al.}(2014)\citenamefont {Bunyk},
  \citenamefont {Hoskinson}, \citenamefont {Johnson}, \citenamefont
  {Tolkacheva}, \citenamefont {Altomare}, \citenamefont {Berkley},
  \citenamefont {Harris}, \citenamefont {Hilton}, \citenamefont {Lanting},
  \citenamefont {Przybysz} \emph {et~al.}}]{bunyk2014architectural}%
  \BibitemOpen
  \bibfield  {author} {\bibinfo {author} {\bibfnamefont {P.}~\bibnamefont
  {Bunyk}}, \bibinfo {author} {\bibfnamefont {E.}~\bibnamefont {Hoskinson}},
  \bibinfo {author} {\bibfnamefont {M.}~\bibnamefont {Johnson}}, \bibinfo
  {author} {\bibfnamefont {E.}~\bibnamefont {Tolkacheva}}, \bibinfo {author}
  {\bibfnamefont {F.}~\bibnamefont {Altomare}}, \bibinfo {author}
  {\bibfnamefont {A.}~\bibnamefont {Berkley}}, \bibinfo {author} {\bibfnamefont
  {R.}~\bibnamefont {Harris}}, \bibinfo {author} {\bibfnamefont
  {J.}~\bibnamefont {Hilton}}, \bibinfo {author} {\bibfnamefont
  {T.}~\bibnamefont {Lanting}}, \bibinfo {author} {\bibfnamefont
  {A.}~\bibnamefont {Przybysz}},  \emph {et~al.},\ }\href@noop {} {\bibfield
  {journal} {\bibinfo  {journal} {IEEE Trans. Appl. Supercond.}\ } (\bibinfo
  {year} {2014})}\BibitemShut {NoStop}%
\bibitem [{\citenamefont {Dickson}(2011)}]{Dickson2011}%
  \BibitemOpen
  \bibfield  {author} {\bibinfo {author} {\bibfnamefont {N.~G.}\ \bibnamefont
  {Dickson}},\ }\href@noop {} {\bibfield  {journal} {\bibinfo  {journal} {New
  J. Phys.}\ }\textbf {\bibinfo {volume} {13}},\ \bibinfo {pages} {073011}
  (\bibinfo {year} {2011})}\BibitemShut {NoStop}%
\bibitem [{\citenamefont {Dickson}\ \emph {et~al.}(2013)\citenamefont {Dickson}
  \emph {et~al.}}]{Dickson2013}%
  \BibitemOpen
  \bibfield  {author} {\bibinfo {author} {\bibfnamefont {N.}~\bibnamefont
  {Dickson}} \emph {et~al.},\ }\href {\doibase 10.1038/ncomms2920} {\bibfield
  {journal} {\bibinfo  {journal} {Nat. Commun.}\ }\textbf {\bibinfo {volume}
  {4}},\ \bibinfo {pages} {1903} (\bibinfo {year} {2013})}\BibitemShut
  {NoStop}%
\bibitem [{\citenamefont {Boixo}\ \emph {et~al.}(2013)\citenamefont {Boixo},
  \citenamefont {Albash}, \citenamefont {Spedalieri}, \citenamefont
  {Chancellor},\ and\ \citenamefont {Lidar}}]{Boixo2013}%
  \BibitemOpen
  \bibfield  {author} {\bibinfo {author} {\bibfnamefont {S.}~\bibnamefont
  {Boixo}}, \bibinfo {author} {\bibfnamefont {T.}~\bibnamefont {Albash}},
  \bibinfo {author} {\bibfnamefont {F.~M.}\ \bibnamefont {Spedalieri}},
  \bibinfo {author} {\bibfnamefont {N.}~\bibnamefont {Chancellor}}, \ and\
  \bibinfo {author} {\bibfnamefont {D.~A.}\ \bibnamefont {Lidar}},\ }\href@noop
  {} {\bibfield  {journal} {\bibinfo  {journal} {Nat. Commun.}\ }\textbf
  {\bibinfo {volume} {4}} (\bibinfo {year} {2013})}\BibitemShut {NoStop}%
\bibitem [{\citenamefont {Albash}\ \emph
  {et~al.}(2015{\natexlab{a}})\citenamefont {Albash}, \citenamefont {Vinci},
  \citenamefont {Mishra}, \citenamefont {Warburton},\ and\ \citenamefont
  {Lidar}}]{Albash2015}%
  \BibitemOpen
  \bibfield  {author} {\bibinfo {author} {\bibfnamefont {T.}~\bibnamefont
  {Albash}}, \bibinfo {author} {\bibfnamefont {W.}~\bibnamefont {Vinci}},
  \bibinfo {author} {\bibfnamefont {A.}~\bibnamefont {Mishra}}, \bibinfo
  {author} {\bibfnamefont {P.~A.}\ \bibnamefont {Warburton}}, \ and\ \bibinfo
  {author} {\bibfnamefont {D.~A.}\ \bibnamefont {Lidar}},\ }\href@noop {}
  {\bibfield  {journal} {\bibinfo  {journal} {Phys. Rev. A}\ }\textbf {\bibinfo
  {volume} {91}},\ \bibinfo {pages} {042314} (\bibinfo {year}
  {2015}{\natexlab{a}})}\BibitemShut {NoStop}%
\bibitem [{Note1()}]{Note1}%
  \BibitemOpen
  \bibinfo {note} {Given $120$ floppy qubits in a bipartite graph such as
  Chimera, there is a {\protect \em stable set} $S$ of at least $60$ qubits
  with no couplings between them; the set of states reached by flipping any
  subset of $S$ is an isoenergetic hypercube of size $2^{|S|}$ in Hamming
  space.}\BibitemShut {Stop}%
\bibitem [{Note2()}]{Note2}%
  \BibitemOpen
  \bibinfo {note} {We avoid degree 2 qubits because in $U_1$ problems they can
  be reduced, as {\protect \em subdivisions} of smaller problems \cite
  {Diestel2012}, and therefore only add complexity in the form of degenerate
  subspaces.}\BibitemShut {Stop}%
\bibitem [{Note3()}]{Note3}%
  \BibitemOpen
  \bibinfo {note} {Instances were generated on distinct random graphs to
  minimize the influence of structural anomalies that might arise in particular
  graphs.}\BibitemShut {Stop}%
\bibitem [{\citenamefont {Amin}(2008)}]{Amin2008}%
  \BibitemOpen
  \bibfield  {author} {\bibinfo {author} {\bibfnamefont {M.~H.~S.}\
  \bibnamefont {Amin}},\ }\href {\doibase 10.1103/PhysRevLett.100.130503}
  {\bibfield  {journal} {\bibinfo  {journal} {Phys. Rev. Lett.}\ }\textbf
  {\bibinfo {volume} {100}},\ \bibinfo {pages} {130503} (\bibinfo {year}
  {2008})}\BibitemShut {NoStop}%
\bibitem [{\citenamefont {Amin}\ and\ \citenamefont {Choi}(2009)}]{Amin2009a}%
  \BibitemOpen
  \bibfield  {author} {\bibinfo {author} {\bibfnamefont {M.~H.~S.}\
  \bibnamefont {Amin}}\ and\ \bibinfo {author} {\bibfnamefont {V.}~\bibnamefont
  {Choi}},\ }\href {\doibase 10.1103/PhysRevA.80.062326} {\bibfield  {journal}
  {\bibinfo  {journal} {Phys. Rev. A}\ }\textbf {\bibinfo {volume} {80}},\
  \bibinfo {pages} {062326} (\bibinfo {year} {2009})}\BibitemShut {NoStop}%
\bibitem [{\citenamefont {Altshuler}\ \emph {et~al.}(2010)\citenamefont
  {Altshuler}, \citenamefont {Krovi},\ and\ \citenamefont
  {Roland}}]{Altshuler2010}%
  \BibitemOpen
  \bibfield  {author} {\bibinfo {author} {\bibfnamefont {B.}~\bibnamefont
  {Altshuler}}, \bibinfo {author} {\bibfnamefont {H.}~\bibnamefont {Krovi}}, \
  and\ \bibinfo {author} {\bibfnamefont {J.}~\bibnamefont {Roland}},\
  }\href@noop {} {\bibfield  {journal} {\bibinfo  {journal} {Proc. Nat. Acad.
  Sci. USA}\ }\textbf {\bibinfo {volume} {107}},\ \bibinfo {pages} {12446}
  (\bibinfo {year} {2010})}\BibitemShut {NoStop}%
\bibitem [{\citenamefont {Young}\ \emph {et~al.}(2010)\citenamefont {Young},
  \citenamefont {Knysh},\ and\ \citenamefont {Smelyanskiy}}]{Young2010}%
  \BibitemOpen
  \bibfield  {author} {\bibinfo {author} {\bibfnamefont {A.~P.}\ \bibnamefont
  {Young}}, \bibinfo {author} {\bibfnamefont {S.}~\bibnamefont {Knysh}}, \ and\
  \bibinfo {author} {\bibfnamefont {V.~N.}\ \bibnamefont {Smelyanskiy}},\
  }\href@noop {} {\bibfield  {journal} {\bibinfo  {journal} {Phys. Rev. Lett.}\
  }\textbf {\bibinfo {volume} {104}},\ \bibinfo {pages} {020502} (\bibinfo
  {year} {2010})}\BibitemShut {NoStop}%
\bibitem [{\citenamefont {Klauder}(1979)}]{Klauder1979}%
  \BibitemOpen
  \bibfield  {author} {\bibinfo {author} {\bibfnamefont {J.}~\bibnamefont
  {Klauder}},\ }\href@noop {} {\bibfield  {journal} {\bibinfo  {journal} {Phys.
  Rev. D}\ }\textbf {\bibinfo {volume} {19}},\ \bibinfo {pages} {2349}
  (\bibinfo {year} {1979})}\BibitemShut {NoStop}%
\bibitem [{\citenamefont {Smolin}\ and\ \citenamefont
  {Smith}(2013)}]{Smolin2013}%
  \BibitemOpen
  \bibfield  {author} {\bibinfo {author} {\bibfnamefont {J.~A.}\ \bibnamefont
  {Smolin}}\ and\ \bibinfo {author} {\bibfnamefont {G.}~\bibnamefont {Smith}},\
  }\href@noop {} {\bibfield  {journal} {\bibinfo  {journal} {arXiv preprint
  arXiv:1305.4904}\ } (\bibinfo {year} {2013})}\BibitemShut {NoStop}%
\bibitem [{\citenamefont {Albash}\ \emph
  {et~al.}(2015{\natexlab{b}})\citenamefont {Albash}, \citenamefont
  {R{\o}nnow}, \citenamefont {Troyer},\ and\ \citenamefont
  {Lidar}}]{Albash2015b}%
  \BibitemOpen
  \bibfield  {author} {\bibinfo {author} {\bibfnamefont {T.}~\bibnamefont
  {Albash}}, \bibinfo {author} {\bibfnamefont {T.~F.}\ \bibnamefont
  {R{\o}nnow}}, \bibinfo {author} {\bibfnamefont {M.}~\bibnamefont {Troyer}}, \
  and\ \bibinfo {author} {\bibfnamefont {D.~A.}\ \bibnamefont {Lidar}},\
  }\href@noop {} {\bibfield  {journal} {\bibinfo  {journal} {Eur. Phys. J.
  Spec. Top.}\ }\textbf {\bibinfo {volume} {224}},\ \bibinfo {pages} {111}
  (\bibinfo {year} {2015}{\natexlab{b}})}\BibitemShut {NoStop}%
\bibitem [{\citenamefont {Shin}\ \emph {et~al.}(2014)\citenamefont {Shin},
  \citenamefont {Smith}, \citenamefont {Smolin},\ and\ \citenamefont
  {Vazirani}}]{Shin2014}%
  \BibitemOpen
  \bibfield  {author} {\bibinfo {author} {\bibfnamefont {S.~W.}\ \bibnamefont
  {Shin}}, \bibinfo {author} {\bibfnamefont {G.}~\bibnamefont {Smith}},
  \bibinfo {author} {\bibfnamefont {J.~A.}\ \bibnamefont {Smolin}}, \ and\
  \bibinfo {author} {\bibfnamefont {U.}~\bibnamefont {Vazirani}},\ }\href@noop
  {} {\bibfield  {journal} {\bibinfo  {journal} {arXiv preprint
  arXiv:1401.7087}\ } (\bibinfo {year} {2014})}\BibitemShut {NoStop}%
\bibitem [{\citenamefont {Muthukrishnan}\ \emph {et~al.}(2015)\citenamefont
  {Muthukrishnan}, \citenamefont {Albash},\ and\ \citenamefont
  {Lidar}}]{Muthukrishnan2015}%
  \BibitemOpen
  \bibfield  {author} {\bibinfo {author} {\bibfnamefont {S.}~\bibnamefont
  {Muthukrishnan}}, \bibinfo {author} {\bibfnamefont {T.}~\bibnamefont
  {Albash}}, \ and\ \bibinfo {author} {\bibfnamefont {D.~A.}\ \bibnamefont
  {Lidar}},\ }\href@noop {} {\enquote {\bibinfo {title} {Tunneling and speedup
  in quantum optimization for permutation-symmetric problems},}\ } (\bibinfo
  {year} {2015}),\ \Eprint {http://arxiv.org/abs/1511.03910v1}
  {arXiv:1511.03910v1} \BibitemShut {NoStop}%
\bibitem [{\citenamefont {Zener}(1932)}]{Zener1932}%
  \BibitemOpen
  \bibfield  {author} {\bibinfo {author} {\bibfnamefont {C.}~\bibnamefont
  {Zener}},\ }in\ \href@noop {} {\emph {\bibinfo {booktitle} {Proc. R. Soc.
  A}}},\ Vol.\ \bibinfo {volume} {137}\ (\bibinfo {organization} {The Royal
  Society},\ \bibinfo {year} {1932})\ pp.\ \bibinfo {pages}
  {696--702}\BibitemShut {NoStop}%
\bibitem [{\citenamefont {King}\ and\ \citenamefont
  {McGeoch}(2014)}]{King2014}%
  \BibitemOpen
  \bibfield  {author} {\bibinfo {author} {\bibfnamefont {A.~D.}\ \bibnamefont
  {King}}\ and\ \bibinfo {author} {\bibfnamefont {C.~C.}\ \bibnamefont
  {McGeoch}},\ }\href@noop {} {\bibfield  {journal} {\bibinfo  {journal} {arXiv
  preprint arXiv:1410.2628}\ } (\bibinfo {year} {2014})}\BibitemShut {NoStop}%
\bibitem [{\citenamefont {Sherrington}\ and\ \citenamefont
  {Kirkpatrick}(1975)}]{Sherrington1975}%
  \BibitemOpen
  \bibfield  {author} {\bibinfo {author} {\bibfnamefont {D.}~\bibnamefont
  {Sherrington}}\ and\ \bibinfo {author} {\bibfnamefont {S.}~\bibnamefont
  {Kirkpatrick}},\ }\href@noop {} {\bibfield  {journal} {\bibinfo  {journal}
  {Phys. Rev. Lett.}\ }\textbf {\bibinfo {volume} {35}},\ \bibinfo {pages}
  {1792} (\bibinfo {year} {1975})}\BibitemShut {NoStop}%
\bibitem [{\citenamefont {Zdeborov{\'a}}\ and\ \citenamefont
  {M{\'e}zard}(2008{\natexlab{b}})}]{Zdeborova2008b}%
  \BibitemOpen
  \bibfield  {author} {\bibinfo {author} {\bibfnamefont {L.}~\bibnamefont
  {Zdeborov{\'a}}}\ and\ \bibinfo {author} {\bibfnamefont {M.}~\bibnamefont
  {M{\'e}zard}},\ }\href {http://stacks.iop.org/1742-5468/2008/i=12/a=P12004}
  {\bibfield  {journal} {\bibinfo  {journal} {J. Stat. Mech. Theor. Exp.}\
  }\textbf {\bibinfo {volume} {2008}},\ \bibinfo {pages} {P12004} (\bibinfo
  {year} {2008}{\natexlab{b}})}\BibitemShut {NoStop}%
\bibitem [{\citenamefont {Zdeborov{\'a}}\ and\ \citenamefont
  {Krzakala}(2011)}]{Zdeborova2011}%
  \BibitemOpen
  \bibfield  {author} {\bibinfo {author} {\bibfnamefont {L.}~\bibnamefont
  {Zdeborov{\'a}}}\ and\ \bibinfo {author} {\bibfnamefont {F.}~\bibnamefont
  {Krzakala}},\ }\href {http://dx.doi.org/10.1137/090750755} {\bibfield
  {journal} {\bibinfo  {journal} {SIAM Journal on Discrete Mathematics}\
  }\textbf {\bibinfo {volume} {25}},\ \bibinfo {pages} {750} (\bibinfo {year}
  {2011})}\BibitemShut {NoStop}%
\bibitem [{\citenamefont {Douglass}\ \emph {et~al.}(2015)\citenamefont
  {Douglass}, \citenamefont {King},\ and\ \citenamefont
  {Raymond}}]{Douglass2015}%
  \BibitemOpen
  \bibfield  {author} {\bibinfo {author} {\bibfnamefont {A.}~\bibnamefont
  {Douglass}}, \bibinfo {author} {\bibfnamefont {A.~D.}\ \bibnamefont {King}},
  \ and\ \bibinfo {author} {\bibfnamefont {J.}~\bibnamefont {Raymond}},\ }in\
  \href@noop {} {\emph {\bibinfo {booktitle} {Theory and Applications of
  Satisfiability Testing--SAT 2015}}}\ (\bibinfo  {publisher} {Springer},\
  \bibinfo {year} {2015})\BibitemShut {NoStop}%
\bibitem [{\citenamefont {Zhuang}(2014)}]{Zhuang2014}%
  \BibitemOpen
  \bibfield  {author} {\bibinfo {author} {\bibfnamefont {Q.}~\bibnamefont
  {Zhuang}},\ }\href {\doibase 10.1103/PhysRevA.90.052317} {\bibfield
  {journal} {\bibinfo  {journal} {Phys. Rev. A}\ }\textbf {\bibinfo {volume}
  {90}},\ \bibinfo {pages} {052317} (\bibinfo {year} {2014})}\BibitemShut
  {NoStop}%
\bibitem [{\citenamefont {Dickson}\ and\ \citenamefont
  {Amin}(2012)}]{Dickson2012}%
  \BibitemOpen
  \bibfield  {author} {\bibinfo {author} {\bibfnamefont {N.~G.}\ \bibnamefont
  {Dickson}}\ and\ \bibinfo {author} {\bibfnamefont {M.~H.}\ \bibnamefont
  {Amin}},\ }\href@noop {} {\bibfield  {journal} {\bibinfo  {journal} {Phys.
  Rev. A}\ }\textbf {\bibinfo {volume} {85}},\ \bibinfo {pages} {032303}
  (\bibinfo {year} {2012})}\BibitemShut {NoStop}%
\bibitem [{\citenamefont {Farhi}\ \emph {et~al.}(2011)\citenamefont {Farhi},
  \citenamefont {Goldstone}, \citenamefont {Gosset}, \citenamefont {Gutmann},
  \citenamefont {Meyer},\ and\ \citenamefont {Shor}}]{Farhi2011}%
  \BibitemOpen
  \bibfield  {author} {\bibinfo {author} {\bibfnamefont {E.}~\bibnamefont
  {Farhi}}, \bibinfo {author} {\bibfnamefont {J.}~\bibnamefont {Goldstone}},
  \bibinfo {author} {\bibfnamefont {D.}~\bibnamefont {Gosset}}, \bibinfo
  {author} {\bibfnamefont {S.}~\bibnamefont {Gutmann}}, \bibinfo {author}
  {\bibfnamefont {H.~B.}\ \bibnamefont {Meyer}}, \ and\ \bibinfo {author}
  {\bibfnamefont {P.}~\bibnamefont {Shor}},\ }\href@noop {} {\bibfield
  {journal} {\bibinfo  {journal} {Quantum Information and Computation}\
  }\textbf {\bibinfo {volume} {11}},\ \bibinfo {pages} {181} (\bibinfo {year}
  {2011})}\BibitemShut {NoStop}%
\bibitem [{\citenamefont {Diestel}(2012)}]{Diestel2012}%
  \BibitemOpen
  \bibfield  {author} {\bibinfo {author} {\bibfnamefont {R.}~\bibnamefont
  {Diestel}},\ }\href@noop {} {\emph {\bibinfo {title} {Graph Theory, 4th
  Edition}}},\ \bibinfo {series} {Graduate texts in mathematics}, Vol.\
  \bibinfo {volume} {173}\ (\bibinfo  {publisher} {Springer},\ \bibinfo {year}
  {2012})\BibitemShut {NoStop}%
\bibitem [{\citenamefont {Isakov}\ \emph {et~al.}(2015)\citenamefont {Isakov},
  \citenamefont {Zintchenko}, \citenamefont {R{\o}nnow},\ and\ \citenamefont
  {Troyer}}]{Isakov2015}%
  \BibitemOpen
  \bibfield  {author} {\bibinfo {author} {\bibfnamefont {S.~V.}\ \bibnamefont
  {Isakov}}, \bibinfo {author} {\bibfnamefont {I.~N.}\ \bibnamefont
  {Zintchenko}}, \bibinfo {author} {\bibfnamefont {T.~F.}\ \bibnamefont
  {R{\o}nnow}}, \ and\ \bibinfo {author} {\bibfnamefont {M.}~\bibnamefont
  {Troyer}},\ }\href@noop {} {\bibfield  {journal} {\bibinfo  {journal}
  {Computer Physics Communications}\ }\textbf {\bibinfo {volume} {192}},\
  \bibinfo {pages} {265} (\bibinfo {year} {2015})}\BibitemShut {NoStop}%
\bibitem [{\citenamefont {Selby}(2014)}]{Selby2014}%
  \BibitemOpen
  \bibfield  {author} {\bibinfo {author} {\bibfnamefont {A.}~\bibnamefont
  {Selby}},\ }\href@noop {} {\bibfield  {journal} {\bibinfo  {journal} {arXiv
  preprint arXiv:1409.3934v1}\ } (\bibinfo {year} {2014})}\BibitemShut
  {NoStop}%
\end{thebibliography}%

\appendix

\section{Experimental details}

\subsection{Quantum annealing processor}

\begin{figure}
\begin{center}
\setlength{\figurewidth}{6.5cm}%
\setlength{\figureheight}{3.7cm}%
\includegraphics{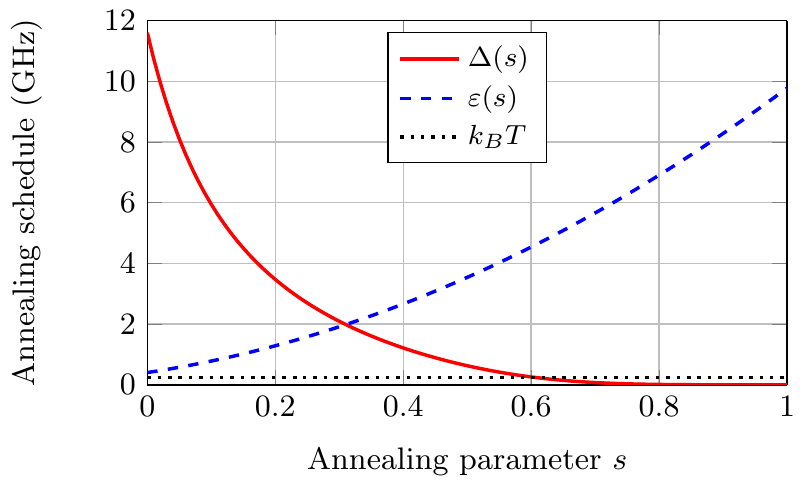}
\caption{Annealing schedule of the D-Wave 2X (DW2X) processor used in this work.  $\Delta(s)/2$ and $\epsilon(s)/2$ represent the prefactors on the transverse and longitudinal Hamiltonians, respectively.  Operating temperature of $12$mK is shown.\label{fig:schedule}}
\end{center}
\end{figure}

The quantum annealing processor used in this work, and in Refs.\ \cite{ttt,Douglass2015}, was a D-Wave Two X processor operating at $12$mK.  Fig.\ \ref{fig:schedule} shows the annealing schedule of the transverse and Ising Hamiltonians (see Eq.\ \ref{eq:ham}).  All experiments were run using an anneal time of $20\si{\micro\second}$, and samples were drawn in batches of 1000 using random spin reversal transformations except where otherwise specified.  Experiments were terminated after finding $100$ ground state samples or drawing $10^5$ samples, except where indicated otherwise (e.g.\ Fig.\ \ref{fig:gauge1}).

\subsection{Simulated annealing}

Simulated annealing was run with a linear schedule in $\beta$ running from $\beta_0=0.01$ to $\beta_f=5$, with the input Hamiltonian $J$ scaled such that $\max_{ij}|J_{ij}|=1$.  For performance scaling experiments shown in Fig.\ \ref{fig:scaling1} performance was measured at optimal anneal length (number of Monte Carlo sweeps) chosen from the set $\{2^3,2^4,\ldots\}$, similar to Ref.\ \cite{flsat}.  In order to estimate the time required in Fig.\ \ref{fig:scaling1} we used the single-thread speed of 6.65 spin updates per nanosecond claimed in Ref.\ \cite{Isakov2015}. Samples were drawn in batches of $100$, and experiments were terminated after finding $100$ ground states or drawing $10^4$ samples.

\section{Mean-field crossing time}

\newcommand{\heffsv}{h_i^{{\rm eff}}(\vec\theta)}

In the context of $\fH_{{\rm SV}}$ (Eq.\ \ref{eq:sssvham}) we consider the effective field $\heffsv$ on qubit $i$ in the state $\vec\theta$ to be
\begin{equation}\label{eq:sssvheff}
\heffsv = h_i + \sum_{j\neq i}J_{ij}\cos\theta_j.
\end{equation}
For a given annealing parameter $s$, consider the {\em locally optimal} value of $\theta_i$ minimizing $\fH_{{\rm SV}}$ subject to $(\theta_j \mid j \neq i)$, i.e.
\begin{equation}\begin{split}
\theta^*_i  =& \argmin_{\theta}\left(\frac{\epsilon(s)}{2} \heffsv \cos\theta- \frac{\Delta(s)}{2}\sin\theta  \right)\\
=& \arccot\left(\frac{-\epsilon(s)\heffsv}{\Delta(s)} \right)
\end{split}
\end{equation}
given that $\Delta(s)$ and $\epsilon(s)$ are positive.  Since modifying $\theta_i$ can alter some $\theta^*_j$, we calculate the associated local minimum $L_s(\vec\theta)$ using an iterative method that replaces $\theta_i$ with $(\theta_i + \theta_i^*)/2$ for every spin $\theta_i$ in turn, and iterates until convergence.

\begin{figure}
\setlength{\figurewidth}{6.5cm}%
\setlength{\figureheight}{3.5cm}%
 
\flushleft\includegraphics{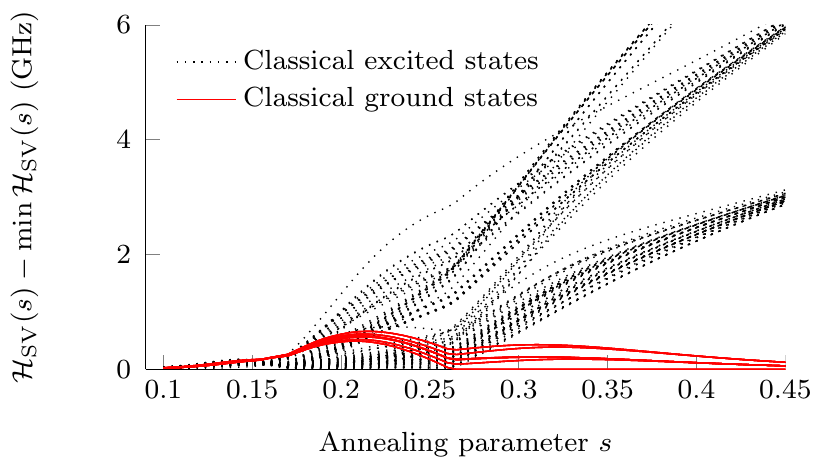}\\\vspace{-.3cm}
\flushleft\includegraphics{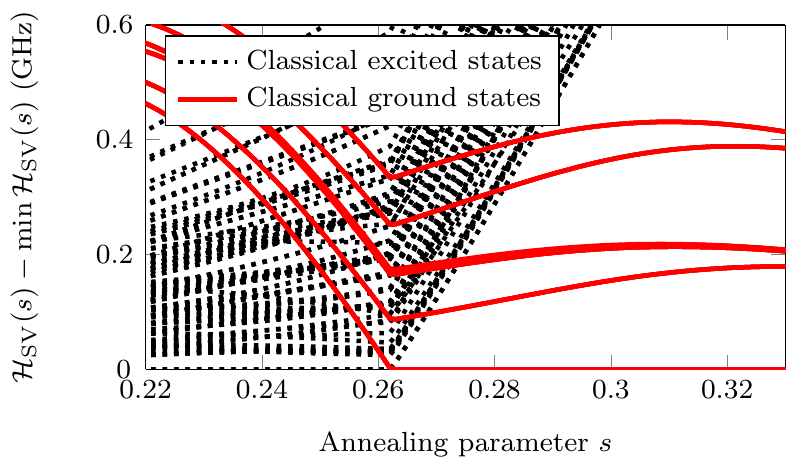}\\
\caption{Mean field spectrum of a $\fC_4$ $U_1^5$ instance with crossing time $s^*_{{\rm SV}}=0.262$.  Excitation of $L_t(\vec\theta)$ from instantaneous spin-vector ground state energy is shown for choices of $\vec\theta$ representing a selection of classical ground and excited states.\label{fig:svspec}}
\end{figure}

Fig.\ \ref{fig:svspec} gives an example mean-field spectrum of a $U_1^5$ instance.  Early in the anneal, $\Delta(s)\gg \epsilon(s)$ and all classical states map to instantaneous ground states in the direction of the transverse field.  Late in the anneal, classical states assume instantaneous mean-field energies that reflect their energies in the final classical Hamiltonian.  From $s^*_{{\rm SV}}=0.262$ onwards, the instantaneous ground state is the associated mean-field local minimum of a classical ground state.

\section{Testbed construction\label{app:testbed}}

Every $U_k^d$ is constructed over a subgraph $G'$ of the available qubit connectivity graph $G$.  For each instance, we construct $G'$ iteratively at random by removing one edge at a time.  An edge $uv$ can be chosen for deletion if:
\begin{enumerate}
\item $u$ has maximum degree in $G'$
\item There is no edge $u'v'$ in $G'$ such that $d(u')\geq d(u)$ and $d(v')>d(v)$.
\item If $u$ and $v$ are in different unit cells (see \cite{bunyk2014architectural}) then $d(u)=d(v)$ and there is no edge $u'v'$ for which both $d(u')=d(v')=d(u)$ and $u'$ and $v'$ are in the same unit cell.
\item $d(u)\geq d(v)\geq 3$.
\item Removing $uv$ does not disconnect the graph.
\end{enumerate}
Given the set of removable edges at a given point in the construction, we choose an edge uniformly at random to remove, and restart the process if we are unable to proceed (i.e.\ reduce the maximum degree of $G'$ to $d$).  Table \ref{tab:degreedist} shows the distributions of degrees for $U_1^d$ instances of varying size.  The distributions are similar for $U_4^d$ instances, as the construction of the graph $G'$ is identical.

\begin{table}
\caption{Degree distributions of instances with target degree $3$ to $6$.  Proportions of qubits with each degree are shown for instance sizes $\fC_4$, $\fC_6$, $\fC_8$ and $\fC_{10}$, which have 128, 288, 512 and 800 qubits respectively (some are inoperable).  Remaining qubits have degree $0$.\label{tab:degreedist}}
\begin{tabular}{c|c c c c c}
$\fC_4$ & degree 2& degree 3& degree 4& degree 5& degree 6\\\hline
$U_1^3$ & 7\% & 92\% & -- &--  &-- \\
$U_1^4$ & --& 4\% & 95\%&-- & --\\
$U_1^5$ & --& --& 4\% &95\% &-- \\
\vspace{1em}$U_1^6$ & --& --&--  &53\% &46\% \\
$\fC_6$ & degree 2& degree 3& degree 4& degree 5& degree 6\\\hline
$U_1^3$ & 10\% & 86\% & -- &--  &-- \\
$U_1^4$ & --& 10\% & 86\%&-- & --\\
$U_1^5$ & --& 2\%& 14\% &81\% &-- \\
\vspace{1em}$U_1^6$ & --& 1\%&7\%  &35\% &53\% \\
$\fC_8$ & degree 2& degree 3& degree 4& degree 5& degree 6\\\hline
$U_1^3$ & 9\% & 88\% & -- &--  &-- \\
$U_1^4$ & --& 8\% & 89\%&-- & --\\
$U_1^5$ & --& 1\%& 10\% &87\% &-- \\
\vspace{1em}$U_1^6$ & --& 1\%&2\% &34\% &61\% \\
$\fC_{10}$ & degree 2& degree 3& degree 4& degree 5& degree 6\\\hline
$U_1^3$ & 10\% & 85\% & -- &--  &-- \\
$U_1^4$ & --& 10\% & 85\%&-- & --\\
$U_1^5$ & --& 2\%& 10\% &83\% &-- \\
\vspace{1em}$U_1^6$ & --& 2\%&5\%  &28\% &61\% \\
\end{tabular}
\end{table}

For $\fC_3$ and $\fC_4$ instances, ground states were computed exhaustively.  For larger instances, we assumed the ground state energy to be the lowest energy found by DW2X, SA, or HFS (see Ref.\ \cite{flsat}).  HFS (run using GS-TW3 \cite{Selby2014}) and SA agreed on the ground state energy on all instances.

\section{Error sensitivity and hardness}

\begin{figure*}
\setlength{\figurewidth}{2.3cm}%
\setlength{\figureheight}{4cm}
\begin{tabular}{rrrrrr}
  \hspace{-.3cm}\includegraphics{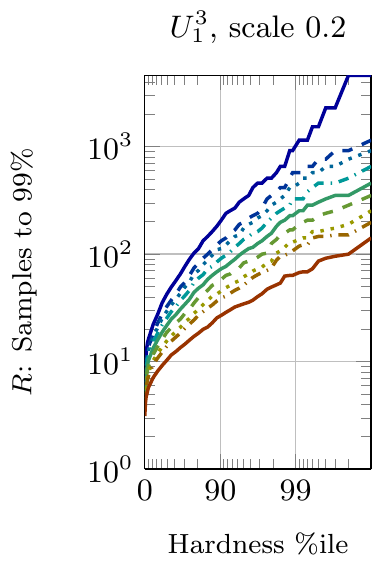}\hspace{-.1cm} &
  \includegraphics{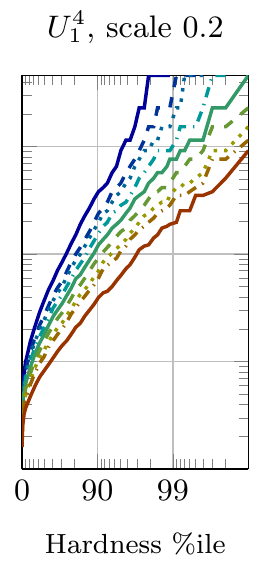}\hspace{.2cm} &
  \includegraphics{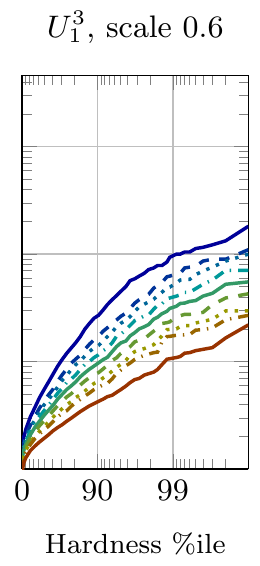}\hspace{-.1cm} &
  \includegraphics{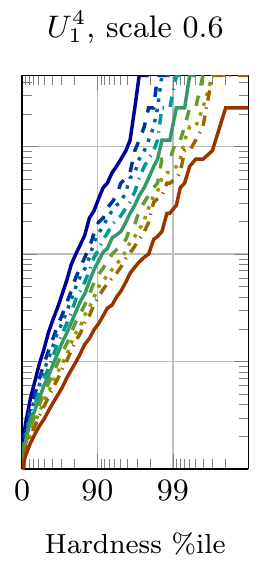}\hspace{.2cm} &
  \includegraphics{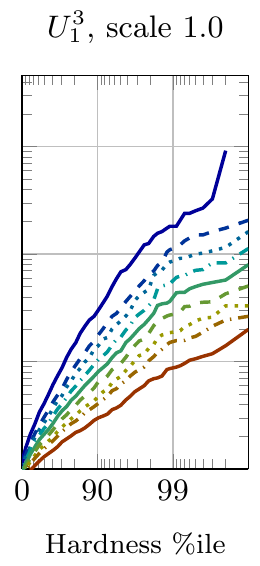}\hspace{-.1cm} &
  \includegraphics{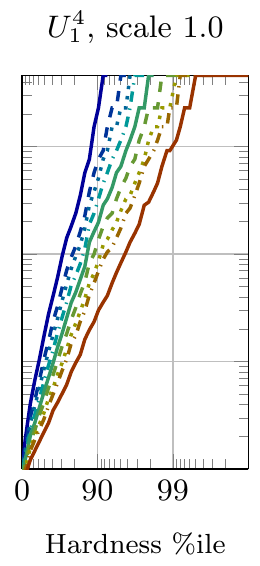}\hspace{-.1cm} \\
  \hspace{-.3cm}\includegraphics{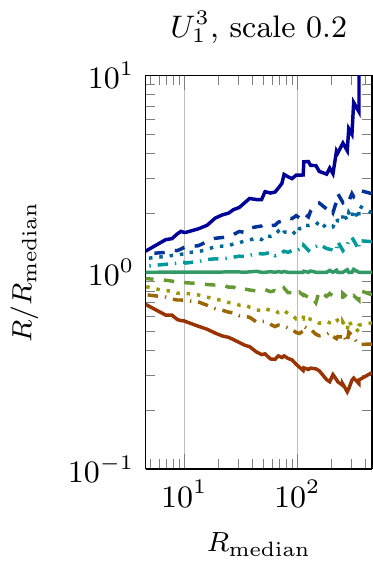}\hspace{0cm} &
  \includegraphics{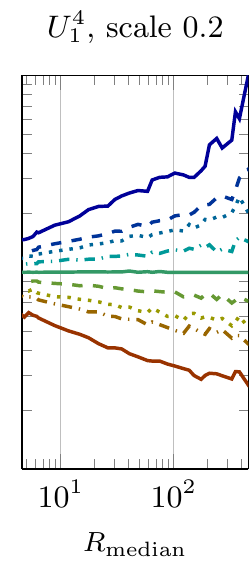}\hspace{.3cm} &
  \includegraphics{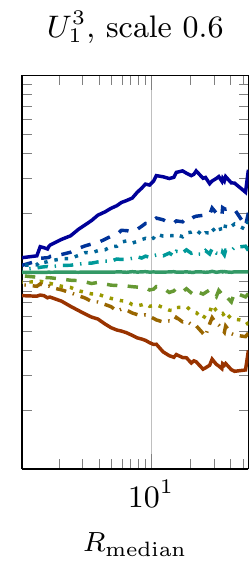}\hspace{.0cm} &
  \includegraphics{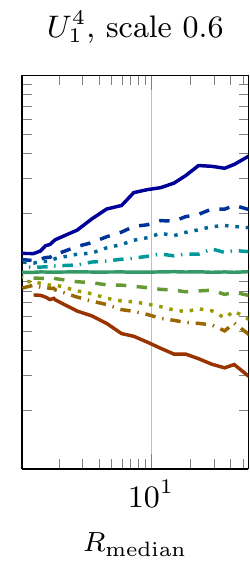}\hspace{.3cm} &
  \includegraphics{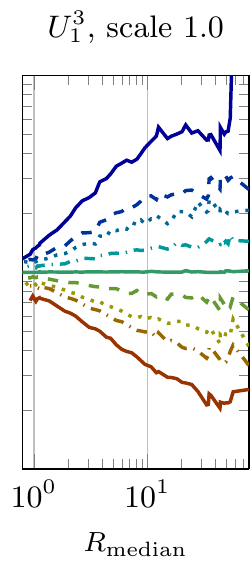}\hspace{.0cm} &
  \includegraphics{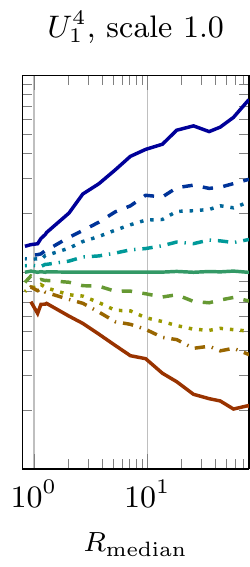}\\
\end{tabular}
\\
Percentile\\
\includegraphics{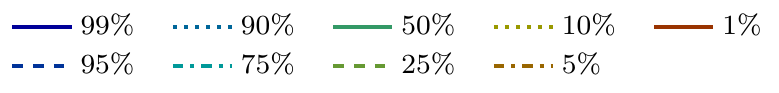}
  \caption{(Top) Varying DW2X performance on $U_1^3$ and $U_1^4$ instances; each instance is run using 10 spin reversal transformations, each of which is repeated 10 times for a total of 100 experiments per instance.  (Bottom) Results for each quantile are divided by median performance $R_{\textrm{median}}$, and plotted against $R_{\textrm{median}}$.  Bottom plots for $U_1^3$ and $U_1^4$ are restricted to the mutual domain.\label{fig:gauge2}}
\end{figure*}

Fig.\ \ref{fig:gauge2} shows error sensitivity data corresponding to Fig.\ \ref{fig:gauge1} when problems are run at energy scale $\alpha\in \{ 0.2,0.6\}$ rather than full energy scale $\alpha=1.0$.  For each energy scale, $U_1^3$ and $U_1^4$ instances of similar hardness $R_{{\rm median}}$ show very similar error sensitivity ($R/R_{{\rm median}}$).

\begin{figure*}
\setlength{\figurewidth}{2.3cm}%
\setlength{\figureheight}{4cm}
\begin{tabular}{rrrrrr}
  \hspace{-.3cm}\includegraphics{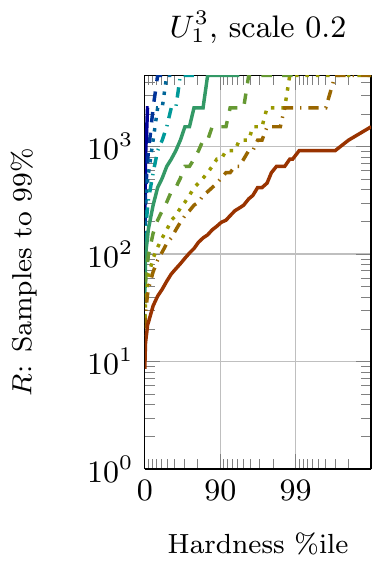}\hspace{-.1cm} &
  \includegraphics{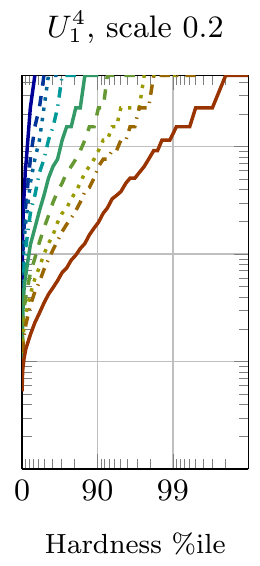}\hspace{.2cm} &
  \includegraphics{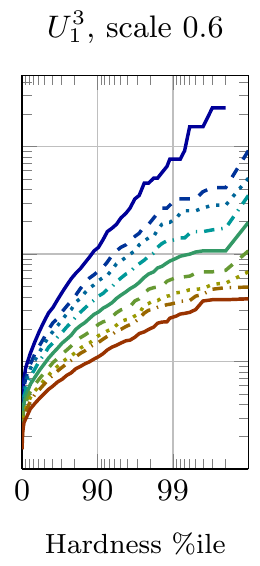}\hspace{-.1cm} &
  \includegraphics{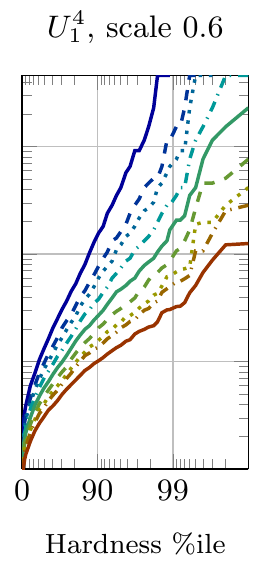}\hspace{.2cm} &
  \includegraphics{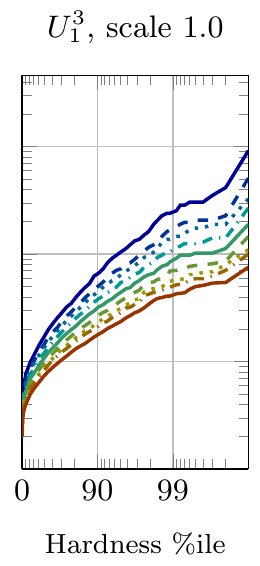}\hspace{-.1cm} &
  \includegraphics{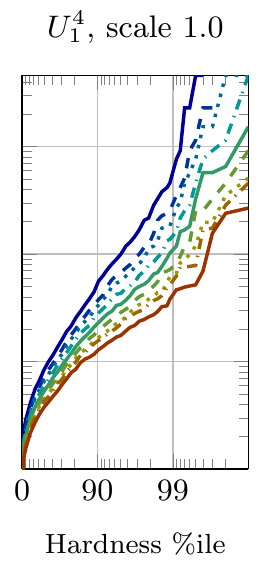}\hspace{-.1cm} \\
  \hspace{-.3cm}\includegraphics{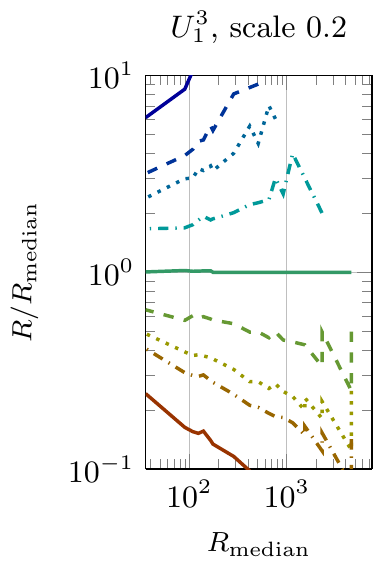}\hspace{0cm} &
  \includegraphics{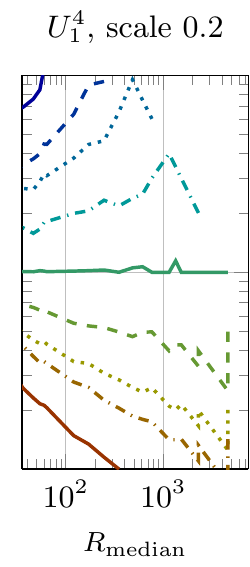}\hspace{.3cm} &
  \includegraphics{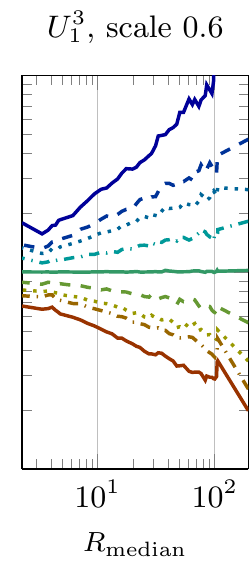}\hspace{.0cm} &
  \includegraphics{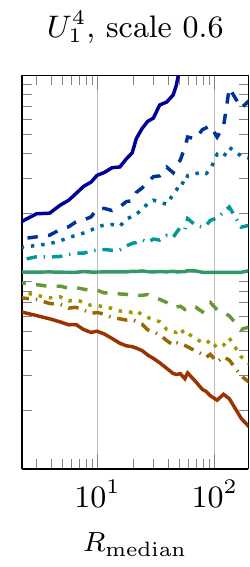}\hspace{.3cm} &
  \includegraphics{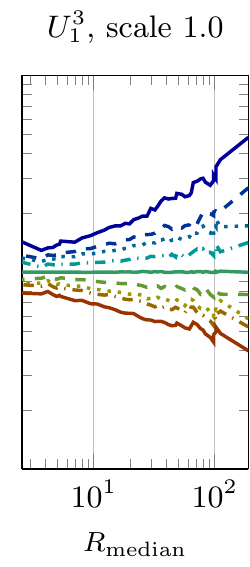}\hspace{.0cm} &
  \includegraphics{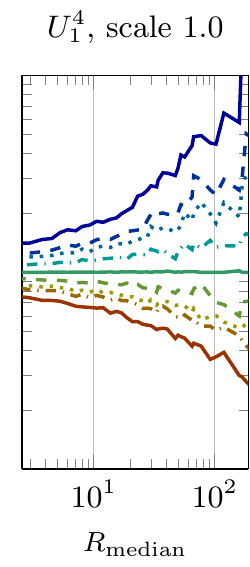}\\
\end{tabular}
\\
Percentile\\
\includegraphics{paper-figure46.pdf}
  \caption{(Top) Varying noisy SA (noise $\sigma = 0.03$) performance on $U_1^3$ and $U_1^4$ instances; each instance is run using 10 spin reversal transformations, each of which is repeated 10 times for a total of 100 experiments per instance.  The experiment quantiles are then sorted by hardness as measured by $R$.  (Bottom) Results for each quantile are divided by the median quantile performance $R_{\textrm{median}}$, and plotted against $R_{\textrm{median}}$.  Bottom plots for $U_1^3$ and $U_1^4$ are restricted to the mutual domain.\label{fig:gaugesa}}
\end{figure*}

Fig.\ \ref{fig:gaugesa} shows analogous data from experiments using {\em noisy SA} as the solver rather than DW2X:  For each experiment consisting of 1000 simulated annealing samples drawn using $2^{10}$ Monte Carlo sweeps, each entry of $h$ and $J$ corresponding to a physically existing coupler or qubit is perturbed by an independent Gaussian error with standard deviation $0.03$.  This error is relative to full energy scale, so experiments run at energy scale $0.2$ will experience Hamiltonian misspecification errors five times as large as those experienced by experiments run at full energy scale.  Both hardness and error sensitivity increase as energy scale decreases, which is both expected due to inflation of relative error, and contrary to what is seen in DW2X data.

\begin{figure*}
\setlength{\figurewidth}{5cm}%
\setlength{\figureheight}{5cm}
\begin{tabular}{rrrr}
  \includegraphics{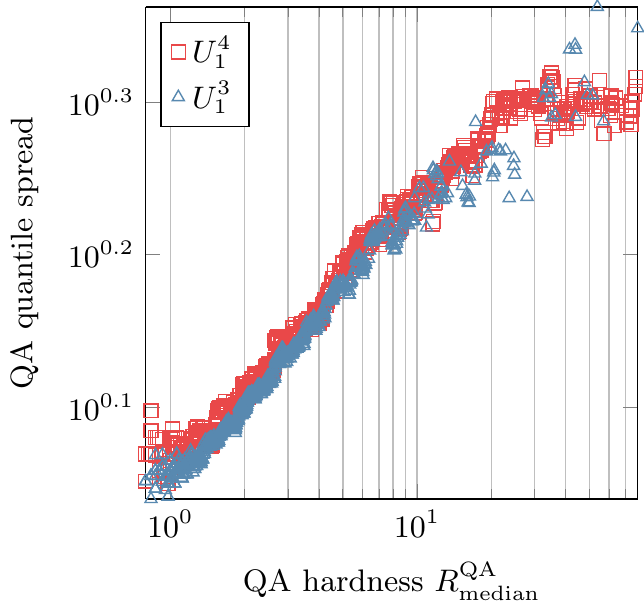}&
  \hspace{2cm}\includegraphics{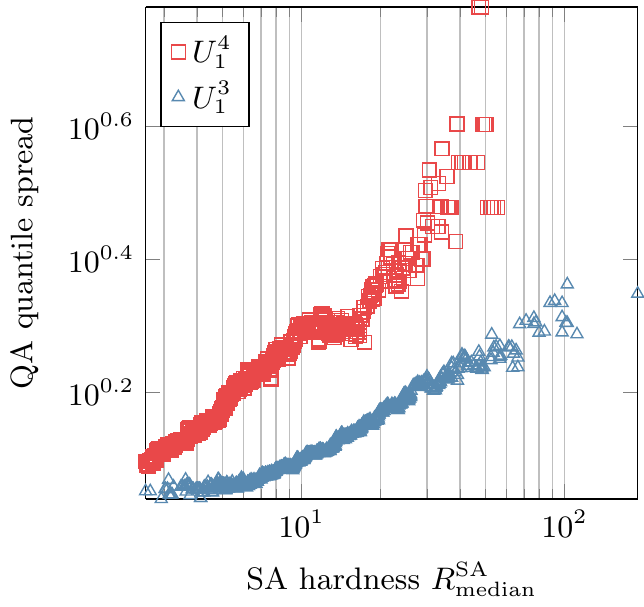}\vspace{1cm}\\
  \includegraphics{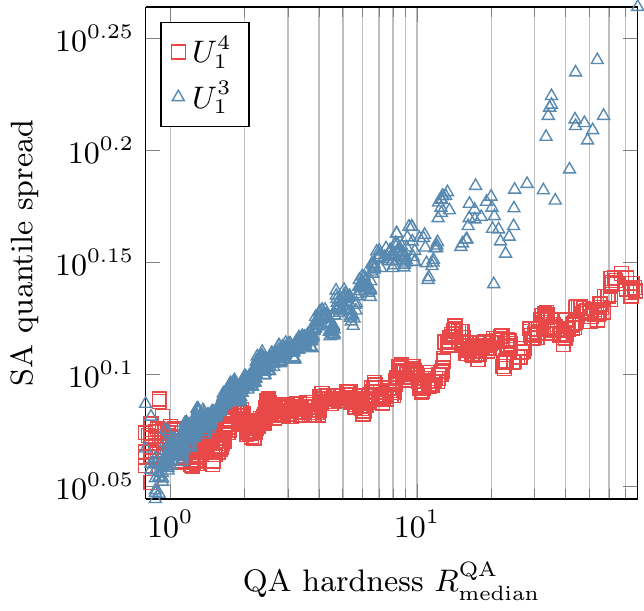}&
  \hspace{2cm}\includegraphics{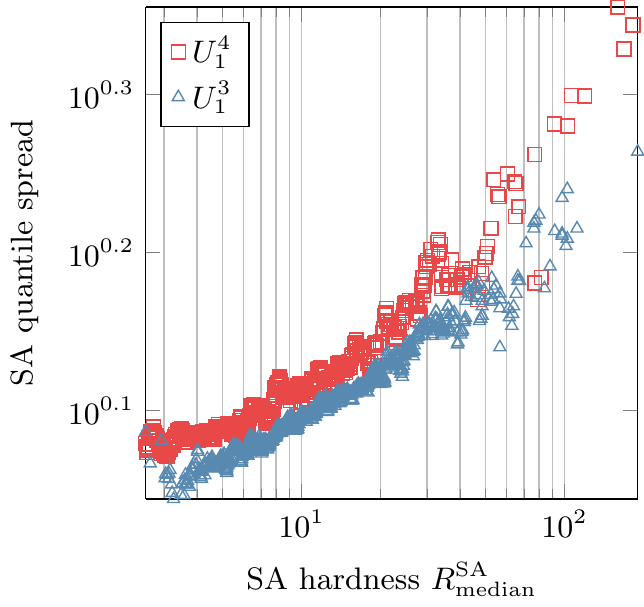}&
\end{tabular}
\caption{Hardness, as seen by QA (DW2X) and noisy SA, versus error sensitivity, as seen by QA and noisy SA.  As in Figs.\ \ref{fig:gauge1} and \ref{fig:gauge2}, we run 100 experiments on each of 1000 $U_1^3$ instances and 1000 $U_1^4$ instances at the $\fC_4$ scale.  We then sort each of the 100 quantiles by hardness, use the entrywise ratio of the 75th and 25th percentiles (the {\em spread} of quantiles) as a measure of error sensitivity, and plot against the median.  Close agreement between $U_1^3$ and $U_1^4$ data indicates a close relationship between error sensitivity as seen by a particular solver, and hardness as seen by a particular solver, regardless of input parameters.  Here we see very close agreement between DW2X error sensitivity and DW2X hardness, and agreement between noisy SA error sensitivity and noisy SA hardness.
\label{fig:spreads}}
\end{figure*}

As further evidence of the relation between DW2X error sensitivity and DW2X hardness, Fig.\ \ref{fig:spreads} plots the spread of relative performance between 75th and 25th percentiles for DW2X and noisy SA, versus hardness as experienced by each solver.  The results indicate that median DW2X performance is a good indicator of DW2X error sensitivity, and that median SA performance is a good indicator of SA error sensitivity, but that median performance of one solver is not as good an indicator of error sensitivity in the other solver.

\section{Performance scaling data}

\begin{figure*}
\setlength{\figurewidth}{3cm}%
\setlength{\figureheight}{4cm}%
\includegraphics{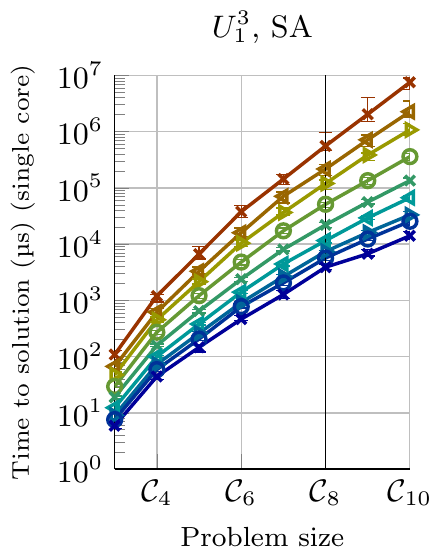}%
\includegraphics{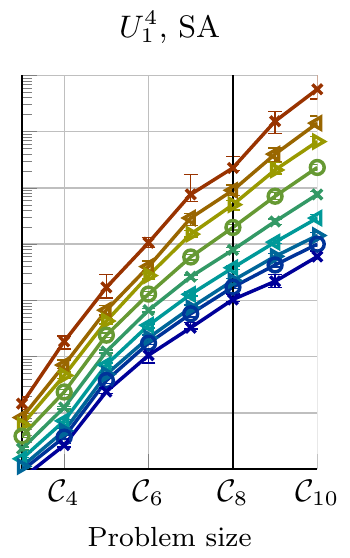}%
\includegraphics{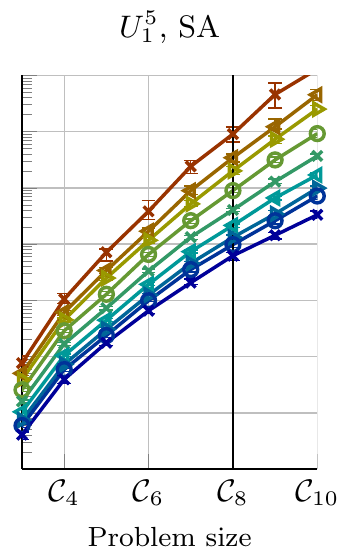}%
\includegraphics{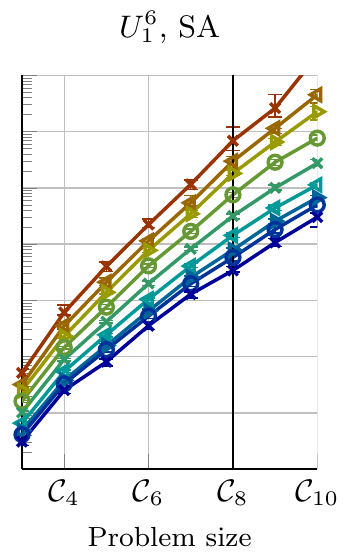}%
\\\vspace{.2cm}
\includegraphics{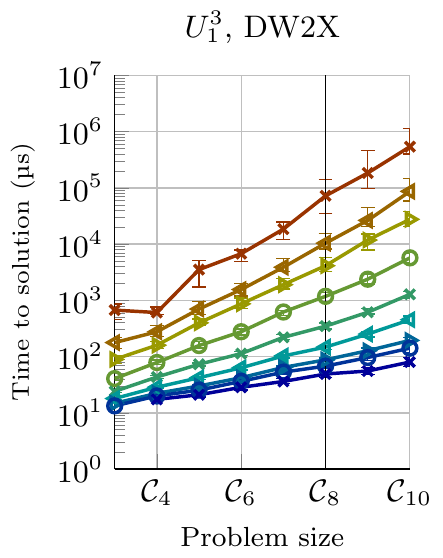}%
\includegraphics{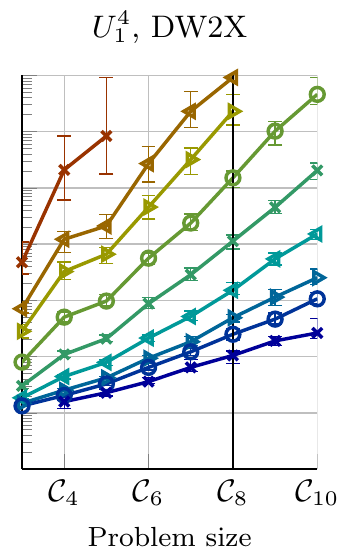}%
\includegraphics{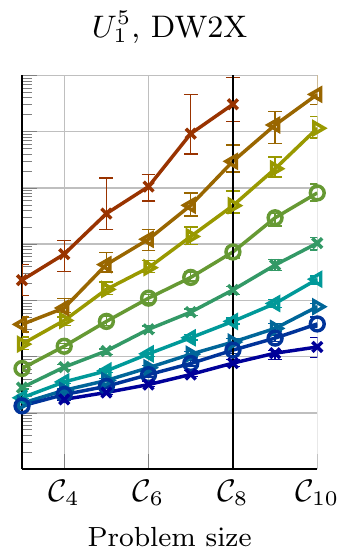}%
\includegraphics{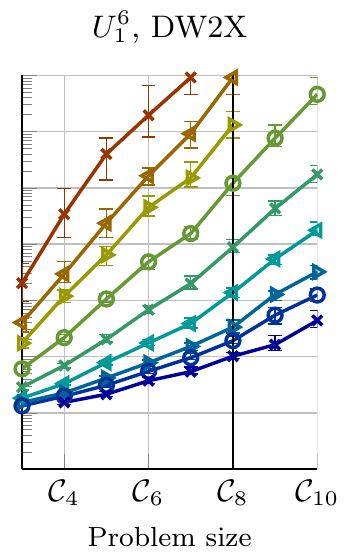}%
\\
Percentile\\
\includegraphics{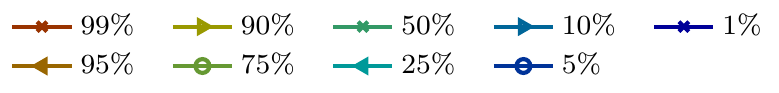}%
\caption{Time to solution for SA (top) and DW2X (bottom) on $U_1^d$ instances with sizes from $\fC_3$ to $\fC_{10}$.  SA uses an optimized number of Monte Carlo sweeps.  DW2X performance on high percentiles indicates the relative absence of heavy tails in $U_1^3$ instances and to a lesser extent $U_1^5$ instances. SA performance indicates that this difference cannot be explained by $U_1^3$ instances being fundamentally easier.\label{fig:scaling1}}
\end{figure*}

Fig.\ \ref{fig:scaling1} shows performance scaling of SA and DW2X on $U_1^d$ instances of varying size for $d\in\{3,4,5,6\}$.  $\fC_L$ instances use up to $8L^2$ qubits in an $L \times L$ subgrid of unit cells in the Chimera graph \cite{ttt}.  At the $\fC_8$ scale, which is the size of the previous generation of D-Wave Two processor, DW2X solves $U_1^3$ problems at the 95th percentile around 1000 times faster than $U_1^4$ problems at the 95th percentile.  As in Ref.\ \cite{Roennow2014}, SA anneal time is optimized, while for DW2X, the minimum available anneal length of $20\si{\micro\second}$ is used; these curves should not be used to characterize the underlying problem-solving dynamics of the DW2X processor.

\end{document}